\documentclass[12pt]{article}
\pdfoutput=1
\usepackage{epsfig}
\usepackage{graphicx}
\usepackage{cite}
\usepackage{amsfonts}
\usepackage{amssymb}
\usepackage{bm}
\usepackage{latexsym}
\usepackage{amsmath}
\numberwithin{equation}{section}
\def\ee{\end{equation}}
\def\ba{\begin{eqnarray}}
\def\ea{\end{eqnarray}}

\def\bq{\begin{quote}}
\def\eq{\end{quote}}

 at 10truept

\newcommand{\order}{{\cal O}}
\newcommand{\beq}{\begin{equation}}
\newcommand{\eeq}{\end{equation}}
\newcommand{\beqa}{\begin{eqnarray}}
\newcommand{\eeqa}{\end{eqnarray}}
\newcommand{\bea}{\begin{eqnarray}}
\newcommand{\eea}{\end{eqnarray}}
\newcommand{\p}{\partial}
\newcommand{\al}{\alpha}
 
 \newcommand{\ep}{\epsilon}

\newcommand{\lle}{\left<}
\newcommand{\rgr}{\right>}
\newcommand{\lb}{\left|}
\newcommand{\rb}{\right|}

\newcommand{\Torder}{\mathrm{T}}
\newcommand{\vect}[1]{\bm{\mathrm{{#1}}}}

\newcommand{\hf}{\frac{1}{2}}
\newcommand{\khat}{{\hat q}}
\newcommand{\lamir}{\Lambda_{IR}}
\newcommand{\dsq}{\frac{d^3q}{(2\pi)^3}}
\newcommand{\zbar}{{\bar \zeta}}

\newcommand{\Hotps}{\left(\frac{H}{2\pi}\right)^2}

\def\ltap{\ \raise.3ex\hbox{$<$\kern-.75em\lower1ex\hbox{$\sim$}}\ }
\def\gtap{\ \raise.3ex\hbox{$>$\kern-.75em\lower1ex\hbox{$\sim$}}\ }
\def\gl{\ \raise.5ex\hbox{$>$}\kern-.8em\lower.5ex\hbox{$<$}\ }
\def\roughly#1{\raise.3ex\hbox{$#1$\kern-.75em\lower1ex\hbox{$\sim$}}}

\begin{document}

\thispagestyle{empty}
\begin{titlepage}
\nopagebreak

\title{
Semiclassical relations and IR effects
in de Sitter\\ and slow-roll space-times} 

\vfill
\author{Steven B. Giddings$^{a}$\footnote{giddings@physics.ucsb.edu}\ \  and Martin S. Sloth$^{b}$\footnote{sloth@cern.ch}
}
\date{ }


\maketitle

\vskip 0.5cm

{\it  $^{a}$ Department of Physics, 
University of California, Santa Barbara, CA 93106}
\vskip 0.5cm
{\it  $^{b}$ CERN, Physics Department, Theory Unit, CH-1211 Geneva 23, Switzerland}

\vfill
\begin{abstract}
We calculate IR divergent graviton one-loop corrections to scalar correlators in de Sitter space, and show that the leading IR contribution may be reproduced via simple semiclassical consistency relations.  One can likewise use such semiclassical relations to calculate leading IR corrections to correlators in slow-roll inflation.  The regulated corrections shift the tensor/scalar ratio and consistency relation of single field inflation, and non-gaussianity parameters averaged over very large distances.  For inflation of sufficient duration, for example arising from a chaotic inflationary scenario, these corrections become of order unity.  First-order corrections of this size indicate a breakdown of the perturbative expansion, and suggest the need for a non-perturbative description of the corresponding regime.  This is analogous to a situation argued to arise in black hole evolution, and to interfere with a sharp perturbative calculation of ``missing information" in Hawking radiation.

 \end{abstract}
 \vskip.4in
 
\noindent
CERN-PH-TH/2010-095\hfill \\  
\vfill
\end{titlepage}

\setcounter{equation}{0} \setcounter{footnote}{0}

\section{Introduction}

While inflation has become a very successful paradigm for early-universe physics, it still leaves us with many challenges and puzzles.  Perhaps the most important of these is incorporating it into a broader framework of quantum gravity, which allows systematic treatment of initial conditions, quantum corrections, {\it etc.}  Even lacking such a treatment, quite a bit of detailed study can be accomplished, treating general relativity as an effective field theory and analyzing it perturbatively.  One of course faces the usual problem of nonrenormalizable UV behavior, but handles this by imagining that appropriate counterterms are provided by a more complete theory, and do not appreciably affect the long distance behavior of the theory relevant for inflationary predictions.  

However, another puzzle arises in the infrared behavior of the theory:  various quantities in de Sitter space, or other inflationary cosmologies, exhibit  IR divergences.  IR divergences typically indicate that one has not understood an important aspect of the dynamics, or has asked a bad question.  For this reason these have been extensively discussed, with arguments ranging from the statement that they can be innocuously eliminated to the statement that they indicate a fundamental instability of de Sitter space causing the cosmological constant to relax to zero\cite{Polyakov:1982ug,Polyakov:2007mm,Polyakov:2009nq,Tsamis:1992sx,Tsamis:1994ca,Tsamis:2007is,Myhrvold:1983hu}.  It is important to more clearly understand where such IR divergences occur, and in what way they modify calculation of physical quantities.

While IR divergent effects of massless scalars have long been known (see {\it e.g.} \cite{Linde:2005ht}), one might question how general these are, given that the correct theory may not have massless scalars.  However, in an effective description of gravity masslessness is inherent in a description in terms of gravitons.  This strongly motivates understanding the IR behavior of processes involving gravitational loops.

Loop calculations in gravity are notoriously complicated, due to the higher spin and its corresponding gauge invariance.  For this reason, so far results have been very limited.  One of the goals of this work is to improve this situation.  In particular, we calculate gravitational loop corrections to propagators in de Sitter space.  While we don't exhibit all such corrections, we note that some of the corrections are indeed IR divergent, and we extract this leading IR behavior.  Moreover, we argue that there is a simple way to derive such behavior from semiclassical arguments, related to the consistency relations of \cite{Maldacena:2002vr,Creminelli:2004yq,Seery:2008ax} (and with antecedents in \cite{Starobinsky:1994bd,Tsamis:2005hd}).  We explicitly check examples where these semiclassical relations match the detailed in-in loop calculations. 

We then propose extending the semiclassical methods to more general, slow-roll, cosmologies, and show how they give leading IR corrections there.  We also demonstrate another check of these methods based on an explicit calculation.  Interestingly, we find that the corrections implied by these methods shift various cosmological observables, notably the tensor/scalar ratio and nongaussianity parameters in the ``large box" (total inflated volume), and we exhibit  leading contributions to such shifts.  Since the corrections grow in the IR, for a sufficiently long period of inflation, for example in models with slow-roll after exit from a chaotic phase, the  loop corrections become of order one, compared to the tree-level quantities.  This suggests a breakdown of the perturbation series, associated with the strong IR effects.

This apparently sharpens the question of the physical interpretation of IR growth in our descriptions of inflationary spacetimes.  While there have been various suggestions that IR effects in de Sitter can be resummed, it seems possible that the breakdown of the perturbation series represents the need for new non-trivial non-perturbative dynamics. 

Indeed, another motivation for this work is an interesting parallel, in black hole physics.  There are good indications that the extreme UV (superplanckian) regime of gravity is governed by physics classically approximated as black hole formation, providing a link between UV and IR physics.  A critical question, in light of Hawking's perturbative arguments\cite{Hawking:1974sw} for information loss, is how such physics can be unitary.  Specifically, by a calculation of the state on a slice spanning the exterior and interior of a semiclassical black hole spacetime, one constructs an argument for information loss.  However, arguments were given in \cite{Giddings:2007ie,Giddings:2009ae} that the perturbative state on such a slice ceases to be a good description, by a sufficiently long time, $t\sim RS$, where $R$ and $S$ are the radius and the entropy of the black hole; one such argument was based on apparently large fluctuations in the perturbative state.  Failure of  such a sharp argument for information loss was proposed as a resolution of the information {\it paradox}, leaving an information {\it problem} of determining the non-perturbative gravitational mechanics that completes the description in this regime.
Quantization on such slices in black hole spacetimes is very similar to quantization on spatial slices in inflationary spacetimes.  The calculations of this paper explicitly show perturbative corrections to the latter that become large on the time scale $t\sim RS$, where now $R$ and $S$ are the de Sitter radius and entropy, in close parallel to the black hole story.\footnote{This analogy was in particular pursued in ref.~\cite{Nimatalk,ArkaniHamed:2007ky}, where it was pointed out than an attempt to regulate inflation by having it terminate through slow-roll fails if the termination time exceeds $t\sim RS$.  The mechanism for this failure is the transition to self-reproduction.}
  This deepens the analogy between the situations, and amplifies the arguments for a similar breakdown in the black hole context.

In outline, the next section summarizes in the de Sitter (dS) case the basic semiclassical relations and resulting leading IR effects on propagators, due to tensor loops. Section three then summarizes checks of these results, through explicit loop calculations of in-in correlators, in which the leading IR behavior is extracted.  Section four extends the semiclassical relations to slow-roll spacetimes, accounting for both tensor and scalar loops, and performs another check on the formalism based on results in the literature.  There is also a brief comment on the relation to the $\delta N$ approach.  Section five calculates the leading corrections to cosmological observables, such as the tensor to scalar ratio, and nongaussianity parameters.  It also contains a worked example of scenarios of slow-roll following chaotic inflation, where such corrections become large due to long-time evolution.  Section six closes with further discussion of interpretation of these IR effects, and the parallels with black hole physics.  The appendices contain review of previously known consistency relations, more complete derivation of the semiclassical consistency relations that we use, and more details of the in-in loop calculations.

\section{Semiclassical relations and IR effects in de Sitter}
\label{sec2}

We first consider correlators and their loop corrections in the simple case of de Sitter space, with a free scalar field $\sigma$ of vanishing expectation value coupled to the metric.  Here the background is fully determined by the cosmological constant $\Lambda$.  The action  is
\beq\label{actions}
S=\frac{1}{2}\int\sqrt{-g}\left[R-\p_{\mu}\sigma\p^{\mu}\sigma -2\Lambda\right]~,
\eeq
where $R$ is the spacetime Ricci scalar, and we choose units $8\pi G = 1/M_p^2=1$.  A perturbative description of the coupling between matter and metric can be derived
using the Arnowitt--Deser--Misner (ADM) formalism
\cite{Arnowitt:1960es},
which gives the action for the coupled perturbations
to any given order by an iterative procedure.
The ADM form of the line element is
\begin{equation}\label{admmetric}
    ds^2= -N^2 dt^2 + h_{ij}(dx^i + N^idt)(dx^j + N^jdt)~,
\end{equation}
where $N$, $N^i$ are the lapse and
the shift functions. The three-dimensional metric $h_{ij}$ encodes
scalar and tensor fluctuations in the spatial geometry. When the line element (\ref{admmetric}) is inserted into Eq.~(\ref{actions}), the lapse
and the shift functions act as Lagrange multipliers: the field
equations obtained by extremizing the action in $N$ and $N^i$ give
the constraint part of the Einstein equations.

The dynamical degrees of freedom are contained within
$\sigma$ and $h_{ij}$.  In de Sitter space there are no dynamical scalar perturbations of the metric; the ``time translation" symmetry implies that the 
usual scalar curvature perturbation $\zeta$, which we will return to in the inflationary case, can be gauged away.  (In more general inflationary spacetimes, this can combine with other scalars to yield a physical degree of freedom.)  In particular, one can choose a gauge where the spatial metric is  parametrized by
\begin{equation}
    h_{ij}=a^2(t)(e^{\gamma})_{ij}~,
\end{equation}
where
\beq
det(e^\gamma)=1\ .
\eeq
The background equations of motion imply $a(t) = a_0\exp(Ht)$, where $H^2 = \Lambda/3$.

In an inflating universe, a leading order description of fluctuations is in terms of pairs of quanta that are produced at the horizon scale, and then pulled apart by the expansion.  When we take into account self interactions -- either in the matter lagrangian, or due to gravity, modes that are produced earlier can influence the correlators of those produced later.  A detailed calculation of this effect can be done via loop corrections in the in-in formalism (see section \ref{in-in}), but an important point is that the leading IR behavior of this effect can be derived by a simple semiclassical procedure.  The basic idea, which is illustrated in study of the corresponding  semiclassical ``consistency relations"\cite{Maldacena:2002vr,Creminelli:2004yq,Seery:2008ax} (and has antecedents in the work of Starobinsky\cite{Starobinsky:1994bd}, and Tsamis and Woodard \cite{Tsamis:2005hd})  is that the early modes are stretched to long wavelength, and the later modes therefore view them as a constant background.  One can compute the correlators of the late modes in the ``background" of the early modes, and then do the required average over fluctuations of this background to determine the resulting corrections to the correlators.

To illustrate this procedure (for a review of known semiclassical relations and further detailed derivation see appendices A and B) consider the variation of the two-point function, $\left<\sigma_{k_1}\sigma_{k_2}\right>$ (where $k_i$ are comoving momenta), due to the effect of long wavelength graviton tensor modes.  This correlator depends on the metric via the magnitude of $k$.  Specifically, for a constant background metric perturbation $\gamma_{ij}$, we have 
\beq
k^2= k_ik_i \rightarrow k_i (e^{-\gamma})_{ij}k_j = k_ik_i -\gamma_{ij}k_ik_j + \hf \gamma_{il}\gamma_{lj} k_i k_j +\cdots\ ;
\eeq
here we use the convention that repeated lower indices are contracted with $\delta_{ij}$.
We can then expand to quadratic order,
\bea\label{specexp}
\left<\sigma_{k_1}\sigma_{k_2}\right>_{\gamma}&=& \left<\sigma_{k_1}\sigma_{k_2}\right>_0 \\&+&\left.\left(-\gamma_{ij}k_ik_j + \hf \gamma_{il}\gamma_{lj} k_i k_j \right)\frac{\p}{\p k^2}\left<\sigma_{k_1}\sigma_{k_2}\right>\right|_0+\frac{1}{2}\left.\left(\gamma_{ij}k_ik_j\right)^2\left(\frac{\p}{\p k^2}\right)^2\left<\sigma_{k_1}\sigma_{k_2}\right>\right|_0+\dots\nonumber
\eea
where the zero subscript denotes vanishing $\gamma$.
The effect of the early, soft gravitons on the correlator can then be estimated by averaging this expression over all such graviton modes; this can be though of as averaging over a ``large box"  (here we drop the zero subscript)
\bea\label{largeav}
\left< \left<\sigma_{k_1}\sigma_{k_2}\right>_{\gamma}\right> &=& \left<\sigma_{k_1}\sigma_{k_2}\right>\\&+&\frac{1}{2}k_ik_j \langle \gamma_{il}\gamma_{lj}\rangle \frac{\p}{\p k^2} \left<\sigma_{k_1}\sigma_{k_2}\right> +\hf k_ik_jk_kk_l \langle\gamma_{ij}\gamma_{kl}\rangle\left(\frac{\p}{\p k^2}\right)^2\left<\sigma_{k_1}\sigma_{k_2}\right>\ . \nonumber
\eea

To compute the average over tensor modes, we first expand
\begin{equation}\label{gravdecomp}
    \gamma_{ij}(x) = \sum_{s=+,\times} \int
    \frac{d^3 k}{(2\pi)^{3}}\left[
        b^{s}_{\vect{k}}\ep^s_{ij}(\vect{k})\gamma_k(t)
        + b^{s\dagger }_{-\vect{k}}\ep_{ij}^{s*}(-\vect{k})\gamma^{*}_k(t)\right]
        e^{i\vect{k}\cdot\vect{x}}~.
\end{equation}
Here $b^{s}_{\vect{k}}$ is an annihilation operator, corresponding to helicity $s$.
The polarization tensors
$\ep^s_{ij}$ are chosen to satisfy
the transversality and tracelessness conditions
$\ep^s_{ii}(\vect k)=k_i\ep^s_{ij}(\vect k)=0$,
together with a completeness relation obtained by tracing over
spatial indices, $\ep_{ij}^s(\vect{k})\ep_{ij}^{\ast s'}(\vect{k})=2\delta_{ss'}$.  The mode functions are the same as those for a scalar, $U_k$, up to normalization; they are most easily written in terms of the conformal time
\beq
\eta = -1/Ha(t) 
\eeq
as
\beq\label{modefcns}
 \gamma_k (\eta) = \sqrt2 U_k(\eta)=
    \frac{H}{\sqrt{k^3}}(1+ik\eta)e^{-ik\eta}\ .
 \eeq
These give a two point function
\beq\label{gravtpt}
\langle\gamma_{ij}(x)\gamma_{kl}(x)\rangle= \sum_s \int  \frac{d^3 q}{(2\pi)^{3}} \frac{H^2}{q^3}(1+q^2\eta^2)\ep^s_{ij}(\vect q)\ep^{s*}_{kl}(\vect q),
\eeq
and the sum over polarizations is \cite{Weinberg:2008zzc}
\bea\label{polarsum}
\omega_{ij,kl}(\vect q)=\sum_s \ep^s_{ij}(\vect q)\ep^{s*}_{kl}(\vect q) &=& \delta_{ik}\delta_{jl}+ \delta_{il}\delta_{jk}-\delta_{ij}\delta_{kl}\\
&+&\delta_{ij}\khat_k\khat_l + \delta_{kl}\khat_i\khat_j- \delta_{ik}\khat_j\khat_l - \delta_{il}\khat_j\khat_k-\delta_{jk}\khat_i\khat_l-\delta_{jl}\khat_i\khat_k + \khat_i\khat_j\khat_k\khat_l\nonumber
\eea
where $\vect {\khat}$ is the unit vector in the direction $\vect{q}$.  We will also define
\beq
\langle \gamma^2(x)\rangle = \frac{1}{4}\langle \gamma_{ij}(x) \gamma_{ij}(x)\rangle = \int \dsq \frac{H^2}{q^3}(1+q^2\eta^2)= 2\Hotps\int\frac{dq}{q}(1+q^2\eta^2)\ .
\eeq

The two different contributions entering (\ref{largeav}) easily follow from these equations: taking $\theta$ to be the angle between $\vect{q}$ and $\vect{k}$ gives
\beq
\frac{k_ik_j}{k^2}\langle \gamma_{il}\gamma_{lj}\rangle = \frac{H^2}{(2\pi)^2}\int \frac{dq}{q}(1+q^2\eta^2) \int d\theta (2\sin^3\theta) = \frac{4}{3} \langle \gamma^2(x)\rangle
\eeq
and
\beq
\frac{k_ik_j}{k^2}\frac{k_kk_l}{k^2}\langle \gamma_{ij}\gamma_{kl}\rangle = \frac{H^2}{(2\pi)^2}\int \frac{dq}{q}(1+q^2\eta^2) \int d\theta (\sin^5\theta) =\frac{8}{15} \langle \gamma^2(x)\rangle\ .
\eeq
The momentum integrals are IR divergent.  The leading IR dependence can be parameterized by introducing an IR cutoff $\lamir$, and for a given time $t_*$, range up to the value $q\sim a_*H$ corresponding to modes just exiting the horizon scale at that time.  A physical origin for such an IR cutoff arises\cite{Vilenkin:1983xp} if inflation began at a time $t_i$, in which case $q>\lamir=a_iH$.  (We may take modes with $q<\lamir$ to be in a different state that is less singular than the Bunch-Davies vacuum.)  Thus these integrals take the form\footnote{In eq.(\ref{gammaIR}) $a_*H$ appears only as a UV cutoff for our semiclassical relations.}
\beq\label{gammaIR}
\left<\gamma^2(x)\right>_* \approx 2\frac{H^2}{(2\pi)^2} \int^{a_*H}_{a_i H}\frac{dq}{q} = 2\frac{H^2}{(2\pi)^2}\log\left(\frac{a_*} {a_i} \right)= 2\frac{H^3}{(2\pi)^2}(t_*-t_i)=  -2\frac{H^2}{(2\pi)^2}\log( \lamir/a_* H) .
\eeq

The result is that the average (\ref{largeav})  yields
\beq\label{sigmaIR}
\left< \left<\sigma_{k_1}\sigma_{k_2}\right>_{\gamma}\right> =\left\{1+\frac{2}{3}\left<\gamma^2(x)\right>_* \left[ \frac{2}{5} k^4\left(\frac{\p}{\p k^2}\right)^2+k^2\frac{\p}{\p k^2}\right]\right\}    \left<\sigma_{k_1}\sigma_{k_2}\right>\ .
\eeq
where $t_*$ is the time of horizon crossing, given by $k=a_*H$, for the mode in question.
The massless scalar two point (Wightman) function
\beq\label{swight}
 \left<\sigma_{k_1}(\eta_1)\sigma_{k_2}(\eta_2)\right> = (2\pi)^3\delta^3(k_1+k_2) U_{k_1}(\eta_1) U^*_{k_2}(\eta_2)\ ,
\eeq
gives a scale invariant spectrum at late times $\eta_1=\eta_2$, 
\beq
\left<\sigma_{k_1}\sigma_{k_2}\right>\approx (2\pi)^3\delta^3(k_1+k_2)\frac{H^2}{2k^3}\ .
\eeq
Precisely for this spectrum, the leading IR correction  in (\ref{sigmaIR}) cancels.\footnote{This cancellation, which preserves scale invariance, can also be verified to hold for a massless scalar in $D$ dimensions, using $\langle \sigma^2\rangle\propto k^{1-D}$ and the statement $\langle \gamma_{ij} \gamma_{kl}\rangle \propto 2\delta_{ij}\delta_{kl} -(D-1)(\delta_{ik}\delta_{jl}+\delta_{il}\delta_{kj})$.}   However, other correlators have different power dependence.  An example is 
\beq
\langle{\dot\sigma}_{k_1}{\dot\sigma}_{k_2}\rangle \approx (2\pi)^3\delta^3(k_1+k_2)\frac{H^4\eta^4}{2}k\ ,
\eeq
where dot denotes $t$ derivative.  So, the preceding steps applied to this correlator yield a correction
\beq\label{sigdotcorr}
\left< \left<{\dot\sigma}_{k_1}{\dot\sigma}_{k_2}\right>_{\gamma}\right>= \left<{\dot\sigma}_{k_1}{\dot\sigma}_{k_2}\right>\left[1+\frac{4}{15} \left<\gamma^2(x)\right>_*\right]\ .
\eeq
Likewise, when we later treat the case of slow-roll, with a spectrum that is not precisely scale invariant, there will be similar corrections.  Note that these loop corrections to correlators become ${\cal O}(1)$ on a time scale $t\sim 1/H^3\sim R S$ after the start of inflation, where $R$ and $S$ denote the de Sitter radius and entropy, respectively.

Analogous arguments apply to the graviton two-point function, with the result 
\beq\label{tol1}
\left< \left<\gamma_{k_1}\gamma_{k_2}\right>_{\gamma}\right>= \left\{1+\frac{2}{3}\left<\gamma^2(x)\right>_* \left[ \frac{2}{5}k^4 \left(\frac{\p}{\p k^2}\right)^2+k^2\frac{\p}{\p k^2}\right]\right\} \left<\gamma_{k_1}\gamma_{k_2}\right>
\eeq
Scale invariance of the graviton correlator in de Sitter likewise implies a vanishing leading correction in this case, but this can be altered by slow-roll.

\section{In-in calculation of leading one-loop corrections}
\label{in-in}

The semiclassical analysis of the preceding section may seem heuristic, and it is important to check the method via an exact calculation.  What is needed is
the one-loop correction to the two-point functions in the ``large box" using the full Schwinger-Keldysh or {\it in-in} formalism.

As in \cite{Maldacena:2002vr}, this calculation is set up by inserting the the ADM decomposition of (\ref{admmetric}) into the action (\ref{actions}), which gives the lagrangian
\bea\label{Lsigma}
\mathcal{L} &=& \frac{a^3}{2}\left[NR^{(3)} -2N\Lambda+ N^{-1}(E^j_i E^i_j-(E^i_i)^2)\right.\nonumber\\
 & &\left. +N^{-1}(\dot\sigma-N^i\p_i\sigma)^2
-N a^{-2}[e^{-\gamma}]^{ij}\p_i\sigma\p_j\sigma\right]~,
\eea
where $E_{ij} =\hf( \dot h_{ij} -\nabla_iN_j-\nabla_jN_i)$ is the rescaled extrinsic curvature and $R^{(3)}$ is the curvature scalar of the three-metric $h_{ij}$.  Since the lapse $N$ and the shift $N^i$ are lagrange multipliers, they were eliminated by their equations of motion, working order-by-order in perturbations about dS, in \cite{Maldacena:2002vr}.

To lowest order we have $N=1$, $N_i=0$ and obtain the free action of gravitons $\gamma$,
\begin{equation}\label{gamma}
    S_2 = \frac{1}{8}\int d^3 x \, dt \; a^3 \left[ 
    \dot{\gamma}_{ij} \dot{\gamma}_{ij} - a^{-2}\partial_k
    \gamma_{ij} \partial_k \gamma_{ij}\right]\; ,
\end{equation}
 and scalars $\sigma$, 
\begin{equation}\label{sigma}
    S_2 = \frac{1}{2} \int d^3 x \, dt \; a^3 \left[ \dot{\sigma }^2 -
    a^{-2}\partial_i \sigma \partial_i \sigma   \right]\; .
\end{equation}
The free fields derived from these actions satisfy the free equations of motion
and can be decomposed into the mode functions given in (\ref{modefcns}).  
With scalar creation and annihilation
operators $a^\dagger_{\vect{k}}, a_{\vect{k}}$, the 
the scalar mode decomposition is
\begin{equation}
    \sigma (x)    = \int \frac{d^3 k}{(2\pi)^{3}}
    \left[a_{\vect{k}}U_k(t)+a^\dagger_{-\vect{k}}U^*_k(t)\right]
    e^{i\vect{k}\cdot\vect{x}}\ , 
\end{equation}
and the graviton mode decomposition was already given in (\ref{gravdecomp}).

We focus on the scalar correlators $\left<\sigma_{k_1}(\eta_0)\sigma_{k_2}(\eta_0)\right>$ and $\left<\dot\sigma_{k_1}(\eta_0)\dot\sigma_{k_2}(\eta_0)\right>$.  The interactions contributing to their one-loop corrections are obtained from the second line of eq.(\ref{Lsigma}).  At cubic order, we have
\beq\label{lthree}
{\cal L}_3 = \frac{a}{2}\gamma_{ij}\p_i\sigma\p_j\sigma\ .
\eeq
At quartic order, the lagrangian is
\beq\label{lfour}
{\cal L}_4 = -\frac{a}{4}\gamma_{il}\gamma_{lj}\p_i\sigma\p_j\sigma +\cdots,
\eeq
where terms with more derivatives are suppressed.  These terms arise from the order-by-order elimination of the lapse and shift, and are given in \cite{Maldacena:2002vr,Seery:2008ax,Dimastrogiovanni:2008af}.
We find that the leading one-loop IR contributions come only from the interactions (\ref{lthree}), (\ref{lfour}).\footnote{This is also checked in \cite{Dimastrogiovanni:2008af}, although that reference obtains different results for the contributions that we calculate below.}

In the in-in formalism the expectation value of any operator $\mathcal{O}$
(evaluated at time $\eta_0$)
is given by
\begin{equation}
    \label{exp1}
    \lle \Omega \rb \mathcal{O}(\eta_0) \lb \Omega \rgr =\lle
    0\rb
    {\bar \Torder}\left(e^{i\int_{-\infty}^{\eta_0}d\eta H_I}\right) \mathcal{O}(\eta_0) T\left(e^{-i\int_{-\infty}^{\eta_0} d\eta H_I}\right)\lb
    0 \rgr
\end{equation}
where $\lb \Omega \rgr$ is the vacuum of the interacting theory,
$\lb 0\rgr$ is the
vacuum of the free theory,
$\Torder$ and $\bar \Torder$ are time ordering and anti-ordering operators, respectively, and $H_I$ is the
interaction Hamiltonian for time $\eta$.  Two topologies of diagrams give leading IR contributions to scalar two point functions at one loop.  The first arises from expanding (\ref{exp1}) to quadratic order in the cubic interaction
\begin{equation}
    H_{3} = - \frac{1}{2}\int d^3 x \; a^2 \gamma_{ij} \partial_i \sigma
    \partial_j \sigma\;,
    \label{eq:hint}
\end{equation}
and is pictured in fig.~1.  The second comes from expanding (\ref{exp1}) to linear order in the analogous $H_4$, and is drawn in fig.~2.
\begin{figure}[!hbtp] \begin{center}
\includegraphics[width=8cm]{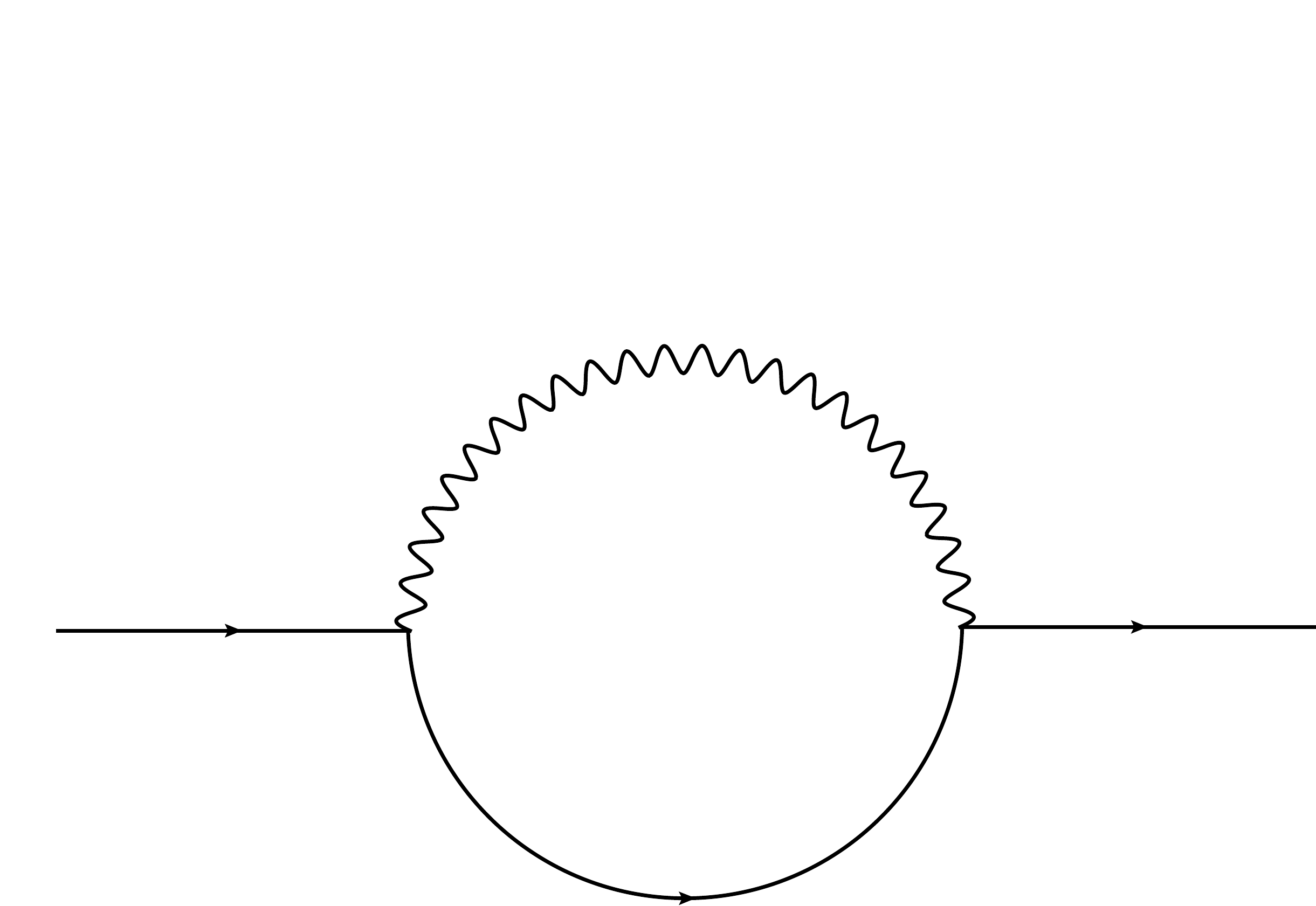}
\end{center}
\caption{Correction to the scalar two point function from an intermediate scalar and graviton.}
\end{figure} 
\begin{figure}[!hbtp] \begin{center}
\includegraphics[width=8cm]{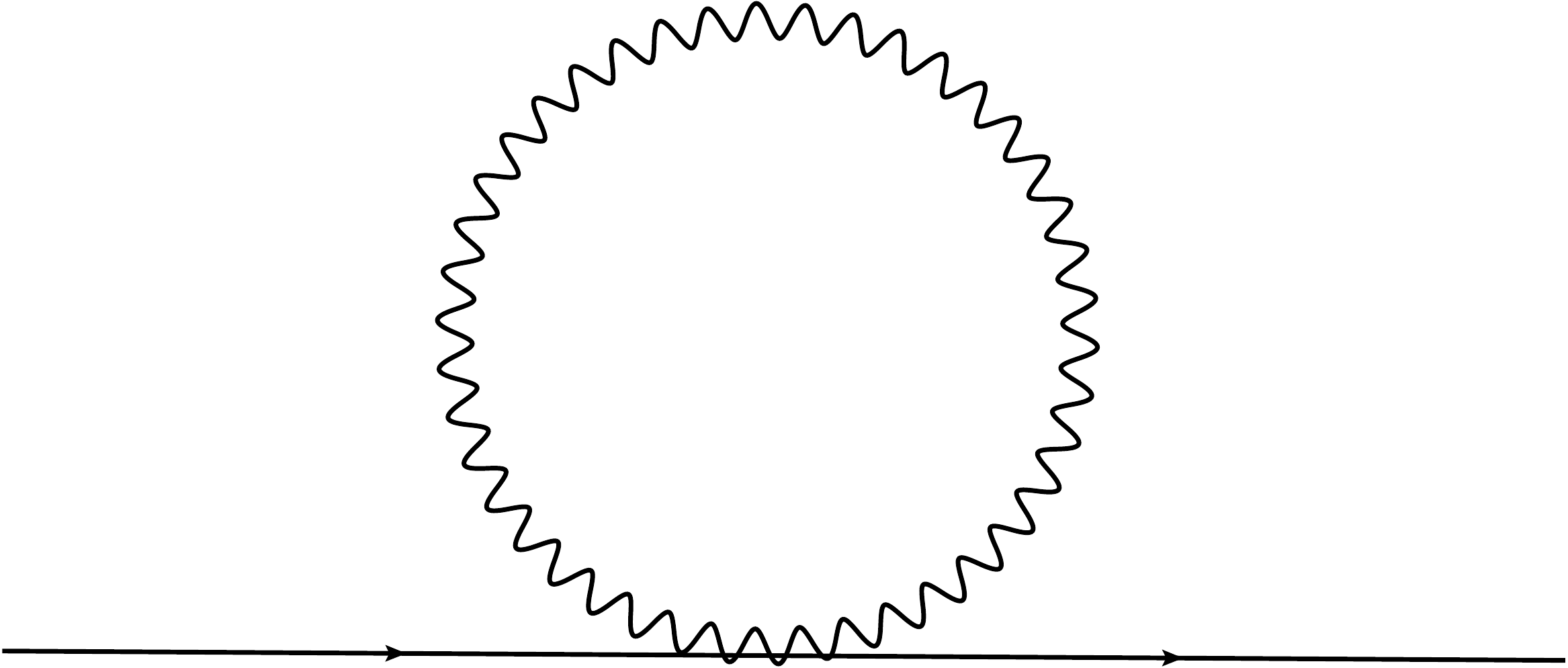}
\end{center}
\caption{Correction to the scalar two point function from a graviton bubble.}
\end{figure}

Two kinds of propagator enter these expressions.  The first are the Wightman functions.  The scalar Wightman function was given in (\ref{swight}),
\beq
 \left<\sigma_{k}(\eta)\sigma_{k'}(\eta')\right> = (2\pi)^3\delta^3(k+k')W_k(\eta,\eta')
 \eeq
 with
 \beq\label{Wscalar}
 W_k(\eta,\eta')=U_{k}(\eta) U^*_{k}(\eta')\ .
 \eeq
 The tensor Wightman function is
\beq
\langle\gamma_{ij}(k,\eta)\gamma_{kl}(k',\eta')\rangle = (2\pi)^3\delta^3(k+k')2\omega_{ij,kl}(k)W_k(\eta,\eta')\eeq
with $\omega_{ij,kl}(k)$ given in (\ref{polarsum}).  One also needs the Feynman functions,
\beq
G_k(\eta,\eta')=\theta(\eta-\eta')W_k(\eta,\eta') + \theta(\eta'-\eta)W_k(\eta',\eta),
\eeq
and correspondingly for the tensors; complex conjugates of these expressions also enter the diagrammatics.

There are two types of contributions from the diagrams of fig.~1, and one from the diagram of fig.~2. Let us parametrize their contributions to the two point function in terms of quantities $A$, $B$, and $C$ respectively, as
\beq
\left<\sigma_{k}(\eta)\sigma_{k'}(\eta)\right> = \left<\sigma_{k}(\eta)\sigma_{k'}(\eta)\right>_0+ (2\pi)^3\delta^3(k+k')[A_k(\eta)+B_k(\eta) + C_k(\eta) +\cdots]\ ,
\eeq
leaving out contributions of other diagrams with more derivatives, which as we have noted are subdominant in the IR.  One can straightforwardly find the first and second terms from the expansion  of (\ref{exp1}) in $H_3$:
\beq
A_k(\eta) = -4{\rm Re}\int \dsq \int_{-\infty}^\eta\prod_{i=1}^2 \frac{d\eta_i}{(H\eta_i)^2}\omega_{ij,kl}(\vect q) k_ik_jk_kk_l G_q(\eta_1,\eta_2) G_{k-q}(\eta_1,\eta_2) W_k(\eta,\eta_1)W_k(\eta,\eta_2)\ ,
\eeq
\beq\label{Bexp}
B_k(\eta)=4{\rm Re}\int \dsq \int_{-\infty}^\eta\prod_{i=1}^2 \frac{d\eta_i}{(H\eta_i)^2}\omega_{ij,kl}(\vect q) k_ik_jk_kk_l W_q(\eta_1,\eta_2) W_{k-q}(\eta_1,\eta_2) W_k(\eta_1,\eta)W_k(\eta,\eta_2)\ .
\eeq
In deriving these expressions we use transversality of $\omega_{ij,kl}(\vect q)$, and can also use $\omega_{ij,kl}(\vect q) k_ik_jk_kk_l=k^4 \sin^4 \theta$, where $\theta$ is the angle between $\vect q$ and $\vect k$.  Likewise, the third term arises from the expansion in $H_4$:
\beq
C_k(\eta)= {\rm Re}\left[ (-2i) \int\dsq \int \frac{d\eta'}{(H\eta')^2} \omega_{il,lj} k_i k_j W_q(\eta',\eta'), W_k(\eta,\eta')^2\right]\ , 
\eeq
in which we can make the replacement $\omega_{il,lj} k_i k_j=2 k^2 \sin^2\theta$.
If we instead want contributions to the two point function of $\dot \sigma_k$, we differentiate both occurrences of $\eta$ in these expressions with respect to $t$. From 
(\ref{Wscalar}) and (\ref{modefcns}), this corresponds to the replacement 
\beq\label{dotreplace}
(1+ik\eta) \rightarrow -H(k\eta)^2
\eeq
 in each of these expressions.

Let us begin with $A_k$, which accounting for the time ordering becomes
\beq
A_k(\eta) = -8{\rm Re}\int \dsq \int_{-\infty}^\eta\frac{d\eta_1}{(H\eta_1)^2} \int_{-\infty}^{\eta_1}\frac{d\eta_2}{(H\eta_2)^2}k^4 \sin^4 \theta
W_q(\eta_1,\eta_2) W_{k-q}(\eta_1,\eta_2) W_k(\eta,\eta_1)W_k(\eta,\eta_2)\ .
\eeq
In this, we substitute (\ref{Wscalar}) and (\ref{modefcns}).  The integrals over $\eta_2$ and $\eta_1$ may then be performed exactly, resulting in a somewhat complicated function of the momenta $\vect k$, $\vect q$, and $\vect p=\vect k-\vect q$.    To do the angular integral, we use
\beq
p^2=k^2+q^2-2kq\cos\theta\ ,
\eeq
giving
\beq
\sin\theta d\theta = \frac{pdp}{kq}\ ,
\eeq
to change variables to an integral over $p$ and $q$, with $p$ ranging from $|k-q|$ to $k+q$.  The integral over $q$, which is UV and IR divergent, may be broken up into the two ranges $q<k$ and $q>k$, with an explicit IR cutoff $\Lambda_{IR}$ used on the lower range, and $\Lambda_{UV}$ bounding the upper range.  (There is no divergence at $q=k$.)  Then, the resulting expression can be expanded in $\Lambda_{IR}$,  $1/\Lambda_{UV}$, and $k\eta$, with small $k\eta$ corresponding to large scales/late times.   

We do not show the intermediate steps, as they are algebraically complicated.  Analogous steps may be performed on the expression (\ref{Bexp}).  Combining these results gives
\beq
A_k(\eta)+B_k(\eta) = \frac{H^2}{2k^3}\frac{H^2}{(2\pi)^2}\left[2\log(k/\Lambda_{IR}) +\frac{2}{3}\log(\Lambda_{UV}/k) + \frac{101}{90} + {\cal O}(k^2\eta^2,\Lambda_{IR}^2,1/\Lambda_{UV}^2)\right]
\eeq
where one can compute the higher-order terms explicitly (see appendix C).
The third integral $C_k$ is more straightforwardly performed, and has expansion
\beq
C_k(\eta) = \frac{H^2}{2k^3}\frac{H^2}{(2\pi)^2}\left[2\log(\Lambda_{IR}/\Lambda_{UV})+ {\cal O}(k^2\eta^2,\Lambda_{IR}^2,1/\Lambda_{UV}^2)\right]\ .
\eeq
Combining these, we find that the IR divergent corrections cancel, as we found in (\ref{sigmaIR}).\footnote{After the first version of this paper appeared, it was shown in \cite{Seery:2010kh} that in the full expression, taking into account other UV divergent terms, the $\log k$ dependence cancels, leaving $\log(\Lambda_{IR}/\Lambda_{UV})$.  Comparing with (\ref{gammaIR}),   $a_*H$ is thus replaced by the true UV cutoff $\Lambda_{UV}$.}  Likewise, repeating the steps for $\langle{\dot\sigma}_{k_1}{\dot\sigma}_{k_2}\rangle$, using the replacement (\ref{dotreplace}) gives
\beq
 \left<{\dot\sigma}_{k_1}{\dot\sigma}_{k_2}\right>\approx \left<{\dot\sigma}_{k_1}{\dot\sigma}_{k_2}\right>_0\left[1- \frac{8}{15}\frac{H^2}{(2\pi)^2} \log(\Lambda_{IR}/k)\right]\ ,
 \eeq
which, using $k=a_*H$ and  (\ref{gammaIR}), is in perfect agreement with (\ref{sigdotcorr}).  

Thus, the results of this section are a non-trivial check of the validity of the semiclassical relations of the preceding section for finding the leading IR corrections to correlators.

\section{Slow-roll}

Having performed non-trivial checks on the validity of the semiclassical relations in de Sitter, we now extend their use to slow-roll inflation. We assume that the methods given in the previous section can be straightforwardly extended to the slow-roll case, and perform various consistency checks on this assumption.
For definiteness and simplicity we study the example of single field slow-roll inflation, although most of the subsequent discussion can be trivially generalized to other models of inflation. 

In single field slow-roll inflation, inflation is driven by a scalar field $\phi$ with a very flat but otherwise arbitrary potential $V(\phi)$,
\beq\label{action}
S=\frac{1}{2}\int\sqrt{-g}\left[R-\p_{\mu}\phi\p^{\mu}\phi -2V(\phi)\right]~,
\eeq
As usual, we define the slow-roll parameters
\beq\label{slowparam}
\epsilon = \hf\left(\frac{V'}{V}\right)^2 \quad,\quad \eta=\frac{V''}{V} \ .
\eeq
Perturbations are again conveniently analyzed using the ADM decomposition (\ref{admmetric}).  Now, however, the scalar degree of freedom mixes non-trivially with the previously gauge scalar degree of freedom of the metric.  Different gauges may be used, but for the following 
considerations we find it useful  to work in the comoving gauge, where the fluctuations in the inflation vanish,
\beq
\phi = \phi_0(t)\ ,
\eeq
with $\phi_0$ the slow-roll solution, and the spatial metric is parameterized as
\begin{equation}
    h_{ij}=a^2(t)e^{2\zeta}(e^{\gamma})_{ij}~.
\end{equation}
Here again $det(e^\gamma)=1$, and in this gauge $\zeta$ can be non-perturbatively identified with the gauge invariant curvature perturbation which is conserved on large scales \cite{Maldacena:2002vr,Lyth:2004gb,Enqvist:2006fs}.

Now, both scalar and tensor fluctuations couple, and contribute to loop corrections.  We consider them in turn.

\subsection{Tensor loops}

The analysis of tensor loop corrections to scalar propagators parallels closely that of section \ref{sec2}.  Specifically, we first begin with 
\bea
\left<\zeta_{k_1}\zeta_{k_2}\right>_{\gamma}&=& \left<\zeta_{k_1}\zeta_{k_2}\right>_0 \\&+&\left.\left(-\gamma_{ij}k_ik_j + \hf \gamma_{il}\gamma_{lj} k_i k_j \right)\frac{\p}{\p k^2}\left<\zeta_{k_1}\zeta_{k_2}\right>\right|_0+\frac{1}{2}\left.\left(\gamma_{ij}k_ik_j\right)^2\left(\frac{\p}{\p k^2}\right)^2\left<\zeta_{k_1}\zeta_{k_2}\right>\right|_0+\dots\nonumber
\eea
We then average this over tensor modes in the ``large box," following the steps of section \ref{sec2}, with the result
\beq\label{zetaIR}
\left< \left<\zeta_{k_1}\zeta_{k_2}\right>_{\gamma}\right> =\left\{1+\frac{2}{3}\left<\gamma^2(x)\right>_* \left[ \frac{2}{5}k^4 \left(\frac{\p}{\p k^2}\right)^2+k^2\frac{\p}{\p k^2}\right]\right\}    \left<\zeta_{k_1}\zeta_{k_2}\right>\ .
\eeq
Now, we have the IR divergent integral
\beq
\left<\gamma^2(x)\right>_* \approx 2 \int^{a_*H}_{a_i H}\frac{dq}{q}\frac{H^2}{(2\pi)^2}\ ,
\eeq
with time-dependent (q-dependent) $H$.  Without scale invariance, the correction doesn't cancel.  If $n_s$ is the scalar spectral index ($n_s=1$ as usual for scale invariance), the expression (\ref{zetaIR}) becomes
\beq\label{sgravcorr}
 \left< \left<\zeta_{k_1}\zeta_{k_2}\right>_{\gamma}\right> =\left[1 + \frac{n_s-4}{3}\frac{n_s-1}{5}  \left<\gamma^2(x)\right>_*\right]\left<\zeta_{k_1}\zeta_{k_2}\right>\ .
 \eeq
 Likewise, for tensors we find
 \beq\label{tgravcorr}
 \left< \left<\gamma_{k_1}\gamma_{k_2}\right>_{\gamma}\right> =\left[1 + \frac{n_t-3}{3}\frac{n_t}{5} \left<\gamma^2(x)\right>_*\right]\left<\gamma_{k_1}\gamma_{k_2}\right>\ ,
 \eeq
where $n_t$ is the tensor index. When applying $(\p/\p k^2)^2$, we will in general also get a contribution from the running of the spectral index $\p n_{s,t}/\p k^2$.  Above we have however neglected the scale dependence of the spectral indices to leading order in slow-roll. More detailed expressions are given in appendix B.

\subsection{Scalar loops}

Accounting for scalar loops follows a similar procedure.  A long-wavelength background scalar $\zbar$ produces the modification
\beq
k^2\rightarrow k^2_\zbar = (e^{-\zbar}k)^2\ .
\eeq
Then, taking into account the dependence of the momentum-space measure on $\zbar$, we have 
\bea\label{zetaexp}
\langle\zeta_{k_1}\zeta_{k_2}\rangle_\zbar &=& \left[1+\zbar \frac{\partial}{\partial \zbar} + \hf \zbar^2\frac{\partial^2}{\partial\zbar^2}+\cdots\right]\left[e^{-6\zbar}\langle\zeta(e^{-\zbar}k_1)\zeta(e^{-\zbar}k_2)\rangle\right]\\
&=&  \left<\zeta_{k_1}\zeta_{k_2}\right>_0-(n_{s}-1)\left.\zbar\left<\zeta_{k_1}\zeta_{k_2}\right>\right|_0+\big(\frac{1}{2}(n_{s}-1)^2+\al_{s}\big)\left.\zbar\zbar\left<\zeta_{k_1}\zeta_{k_2}\right>\right|_0+\dots\nonumber
\eea
Here we have introduced the running of the spectral index,\footnote{In the first eprint version of this paper, we had neglected the running of the spectral indices, but it was subsequently pointed out in \cite{Byrnes:2010yc} that these are generically relevant in the case of scalar loops.} $\al_{s} = d n_s/d\ln(k)$, and we only give expressions to leading order in slow-roll. More complete expressions are given in appendix B.

So, then  again we obtain the effect of the soft scalar mode on the two point function, by averaging over the two-point correlation function in the large box 
\bea\label{sol2}
\left< \left<\zeta_{k_1}\zeta_{k_2}\right>_{\zbar}\right> &\simeq& \left<\zeta_{k_1}\zeta_{k_2}\right>_0+\big(\frac{1}{2}(n_{s}-1)^2+\al_{s}\big)\left<\zeta_{k_1}\zeta_{k_2}\right>_0\left<\zeta^2(x)\right>_*~,
\eea
with 
\beq\label{zvar}
\left<\zeta^2(x)\right>_*\approx  \int^{a_*H}_{a_i H}\frac{dq}{q}\frac{1}{2\epsilon}\frac{H^2}{(2\pi)^2}\ .
\eeq
Similarly, introducing $\al_{t} = d n_t/d\ln(k)$,
\bea\label{sol3}
\left< \left<\gamma_{k_1}\gamma_{k_2}\right>_{\zbar}\right> &\simeq& \left<\gamma_{k_1}\gamma_{k_2}\right>_0+\big(\frac{1}{2}(n_t)^2+\al_{t}\big) \left<\gamma_{k_1}\gamma_{k_2}\right>_0\left<\zeta^2(x)\right>_*~,
\eea
here we use the standard convention where for historical reasons $n_s=1$ corresponds to a scale invariant scalar spectrum, while $n_t=0$ corresponds to a scale invariant tensor spectrum. 
The corrections are suppressed by  slow-roll factors
\beq
n_s-1 =2\eta-6\ep \quad,\quad n_t=-2\epsilon\ ;
\eeq
they vanish in pure de Sitter, where the scalar mode $\zeta$ decouples, as expected.

\subsection{Alternative explicit check}

As a check on the semiclassical relation with scalars, we can use an explicit calculation of the  
second order gauge invariant curvature perturbation\cite{Acquaviva:2002ud}\footnote{The curvature perturbation $\mathcal{R}$ is related to our $\zeta$ by $\mathcal{R}=-\zeta$ as discussed in the second footnote of \cite{Weinberg:2005vy} around eq.(3). The curvature perturbation in \cite{Acquaviva:2002ud} is related to the conserved curvature perturbation $\mathcal{R}$ only in the case of  $\al_s=0$, as was pointed out in \cite{Vernizzi:2004nc}. We thank Filippo Vernizzi for bringing this to our attention.}  $\mathcal{R}^{(2)}$. Ref.~\cite{Acquaviva:2002ud} obtained in the case of  $\al_s=0$
\beq
\mathcal{R}^{(2)} =-\frac{1}{2}(n_s-1)(\mathcal{R}^{(1)})^2 +\mathcal{I}
\eeq
where the non-local contribution $\mathcal{I}$ is determined by the integral 
\bea
 \mathcal{I} &=&-\frac{2}{\ep}\int\frac{1}{a^2}\psi^{(1)}\p_i\p^i\psi^{(1)} dt-\frac{4}{\ep}\int\frac{1}{a^2}(\p_i \psi^{(1)}\p^i\psi^{(1)}) dt\nonumber\\
 & &-\frac{4}{\ep}\int (\ddot{\psi}^{(1)})^2 dt +(\ep-\eta)\Delta^{-1}\p_i\mathcal{R}^{(1)}\p^i\mathcal{R}^{(1)}~,
\eea
and on large scales $\psi^{(1)}=\ep \mathcal{R}^{(1)}$.
The one-loop contribution to $\left< \mathcal{R}_k\mathcal{R}_{k'}\right>$ is then given by $\left<\mathcal{R}^{(2)}_k\mathcal{R}^{(2)}_{k'}\right>$.

First we note that on large scales all higher order gradient terms and time derivatives of $\psi^{(1)}$ vanishes to leading order in slow-roll.  So on large scales to leading order, we have
\beq
 \mathcal{I} \sim (\ep-\eta)\Delta^{-1}\p_i\mathcal{R}^{(1)}\p^i\mathcal{R}^{(1)},
\eeq
which in Fourier space gives
\beq
 \mathcal{I}(k) \sim (\ep-\eta) \int\frac{d^3 q}{(2\pi)^3}\frac{{\bf q}\cdot{\bf k}-q^2}{k^2}\mathcal{R}^{(1)}_{\bf q}\mathcal{R}^{(1)}_{\bf k-q}
\eeq
so the potential IR divergence in $\mathcal{R}^{(1)}_{\bf q}$ from the lower limit of the $q$-integral, is always regulated by a multiplicative power of $q$  in the limit $q\to 0$, and there are no IR divergences from this piece to leading order in slow-roll.

So, let us finally look at the contribution to $\left<\mathcal{R}^{(2)}_k\mathcal{R}^{(2)}_{k'}\right>$ from the local piece. From that we obtain
\bea
\left<\mathcal{R}^{(2)}_k\mathcal{R}^{(2)}_{k'}\right>_{loc} &=& \frac{1}{4}(n_s-1)^2 \int\frac{d^3 q_1}{(2\pi)^3} \int\frac{d^3 q_2}{(2\pi)^3}\left<\mathcal{R}^{(1)}_{{\bf q}_1}\mathcal{R}^{(1)}_{{\bf k-q}_1}\mathcal{R}^{(1)}_{{\bf q}_2}\mathcal{R}^{(1)}_{{\bf k'-q}_2}\right>\nonumber\\
&=&\frac{1}{2}(n_s-1)^2  \int\frac{d^3 q_1}{(2\pi)^3} \int\frac{d^3 q_2}{(2\pi)^3} \left<\mathcal{R}^{(1)}_{\bf q_1}\mathcal{R}^{(1)}_{{\bf q_2}}\right>\left<\mathcal{R}^{(1)}_{\bf k-q_1}\mathcal{R}^{(1)}_{\bf k'-q_2}\right>\ .
\eea
If we split the integral over $q$ into an IR part $a_iH_i < q< k = a_*H_*$ and a UV part $q> a_*H_*$, then from the IR part of the integral, we obtain to leading order in slow-roll
\beq
\left<\mathcal{R}^{(2)}_k\mathcal{R}^{(2)}_{k'}\right>^{IR}_{loc} =\frac{1}{2}(n_s-1)^2 \left<\mathcal{R}^{(1)}_{\bf k}\mathcal{R}^{(1)}_{\bf k'}\right> 
\left< {\mathcal{R}^{(1)}}^2(x)\right>_*~,
\eeq
in agreement with eq.~(\ref{sol2}) in the case $\al_s=0$.

\subsection{Comments on relation to $\delta N$ approach}

In the $\delta N$ approach, the relative curvature perturbation on super-horizon scales can be described by the evolution of causally disconnected regions, which are assumed to evolve like separate unperturbed universes. In this approach, which corresponds to the leading order in the gradient expansion, one has $\zeta =\delta N$, where $N$ is the number of e-folds of expansion between an initially flat spatial hypersurface and a final uniform density surface\footnote{For more details, see in particular the discussion in section IV of \cite{Lyth:2004gb}.} \cite{Starobinsky:1982ee,Starobinsky:1986fxa,Sasaki:1995aw,Lyth:2004gb,Lyth:2005fi}. This implies that in single field slow-roll inflation, one can write the comoving curvature perturbation as an expansion of the field perturbation at horizon crossing \cite{Byrnes:2007tm}
\beq\label{dNtrans}
\zeta = \left.N' \delta\phi\right|_{hor. cros.} +  \frac{1}{2}\left. N''  \delta\phi\star\delta\phi\right|_{hor. cros.} +\dots
\eeq
which now describes the classical relation on super-horizon scales between $\zeta$ and $\delta\phi$, or equivalently it can be viewed as the super-horizon part of the gauge transformation between the two variables in uniform curvature gauge and comoving gauge respectively. Note that in eq.(\ref{dNtrans}) above, prime denotes derivation with respect to the field $\phi$, and star denotes a convolution of momenta in Fourier space. As an example, by use of the chain rule, one can to linear order write $N'\equiv d\ln(a)/d\phi = (dt/d\phi) (d\ln(a)/dt) = H/\dot\phi$, in order to recover the linear relation between $\zeta$ and $\delta\phi$ on large scales.

In this approach one can calculate the one-loop corrections to correlation functions of $\zeta$, by calculating one-loop correlation functions of $\delta\phi$ at horizon crossing, and relating them by the $\delta N$ approach, which is equivalent to the appropriate gauge transformation. In Fourier space, the $\delta N$ transformation in eq.(\ref{dNtrans}) will at the non-linear level introduce new momentum integrals, called $\delta N$ loops in the literature, which can contain additional IR divergences. 

However, it is known that already at horizon crossing the correlation functions of $\delta\phi$ contains IR divergences \cite{Sloth:2006az,Sloth:2006nu}, which cannot be captured by $\delta N$ loops. Thus the IR effects computed from $\delta N$ loops \cite{Byrnes:2007tm,Seery:2007wf,Lyth:2007jh,Bartolo:2007ti,Enqvist:2008kt} are incomplete. One important manifestation of this incompleteness is that the tensor modes do not transform in the $\delta N$ approach (they are gauge invariant on super-horizon scales) and hence there are no $\delta N$ loop corrections for correlation functions of tensor modes, while the calculations of the previous sections clearly show that the correlation functions of tensor modes receive IR one-loop corrections.

\section{Contributions to cosmological observables}

The leading order IR loop corrections that we have found can produce corrections to cosmological observables, which can become significant for sufficiently long periods of inflation.  In this section we first discuss corrections to the tensor/scalar ratio and nongaussianity parameters, averaged over large distances, and then estimate the size of the corrections in a model where slow-roll begins with an exit from the chaotic regime.

\subsection{Tensor-scalar ratio}

The tensor-scalar ratio is one of the main observables in inflationary cosmology. The absence of a detection of primordial tensor modes severely constrains the tensor-scalar relation and leads to an exclusion of $\lambda\phi^4$ inflation (see fig.(14) of \cite{Spergel:2006hy}). In slow-roll inflation there is at tree-level a definite prediction for the tensor-scalar ratio, called the tensor-scalar consistency relation \cite{Wands:2003pw}.  We show that the IR loop corrections that we have found shift this relation in the ``large box".

The tensor-scalar ratio, $r$, is defined by
\beq
r= 4\frac{\left<|\gamma_k|^2\right>}{\left<|\zeta_k|^2\right>}~,
\eeq
where at tree-level in slow-roll we have $\left<|\zeta_k|^2\right>_0 =(1/4\ep) \left<|\gamma_k|^2\right>_0$, which implies the single field tensor-scalar consistency relation (see e.g. eq.(23) of \cite{Wands:2003pw})
\beq
r = 16\ep~.
\eeq

Now, let us consider the one-loop result. 
For the scalar spectrum, we combine (\ref{sol2}) and (\ref{sgravcorr}) to find 
\beq\label{Tscalar}
\left< \left<\zeta_{k_1}\zeta_{k_2}\right>\right> = \left<\zeta_{k_1}\zeta_{k_2}\right>_0\left[1+ \big(\frac{1}{2}(n_{s}-1)^2+\al_s\big)\left<\zeta^2(x)\right>_*+ \frac{n_s-4}{3}\frac{n_s-1}{5} \left<\gamma^2(x)\right>_*     \right]\ .
\eeq
Likewise, for the tensor spectrum, we combine (\ref{sol3}) and (\ref{tgravcorr}): 
\beq\label{Ttensor}
\left< \left<\gamma_{k_1}\gamma_{k_2}\right>\right> = \left<\gamma_{k_1}\gamma_{k_2}\right>_0\left[1+\big(\frac{1}{2}(n_t)^2 +\al_t\big)\left<\zeta^2(x)\right>_*+ \frac{n_t-3}{3}\frac{n_t}{5} \left<\gamma^2(x)\right>_*\right]\ .
\eeq
Using these gives the corrections to the ratio at one-loop order:
\beq
r=16\ep\left[1+\frac{n_t^2+2\alpha_t-((n_s-1)^2+2\alpha_s)}{2} \langle\zeta^2(x)\rangle_* +\frac{(n_t-3)n_t-(n_s-4)(n_s-1)}{15}\langle\gamma^2(x)\rangle_*\ \right]\ .
\eeq
From this we see that the loop corrections we have found generically alter the slow-roll tensor-scalar consistency relation.
So while the loops don't change the scaling of the two-point functions, they do contribute shifts to the averaged tensor/scalar ratio,  as anticipated in \cite{Sloth:2006nu}. In cases where the scalar fluctuations dominate (for one such case, see section \ref{Chaoi}), note that $n_t^2+2\alpha_t>(n_s-1)^2+2\al_s$ would imply $r>16 \epsilon$.

\subsection{Non-Gaussianity}

The non-gaussianity of primordial cosmological perturbations arising from the three point function of curvature perturbations is usually parametrized by the dimensionless parameter $f_{NL}$, which gives the relative strength between the three point function and the two point function.

Just as the power spectrum, $P_{\zeta}(k)$, of curvature perturbations is defined by
\beq
\left<\zeta_{{\bf k}_1}\zeta_{{\bf k}_2}\right> \equiv (2\pi)^3\delta^3({\bf{k}}_1+{\bf{k}}_2)P_{\zeta}(k)~,
\eeq
which is related to the usual scale independent power spectrum by $P_{\zeta}(k)=(2\pi^2)\mathcal{P}_{\zeta}(k)/k^3$, one can similarly one can define the bispectrum  $B_\zeta$ by
\begin{equation}
    \langle \zeta(\vect{k}_1)
        \zeta(\vect{k}_2)
        \zeta(\vect{k}_3) \rangle
    \equiv (2\pi)^3 \delta(\sum_a \vect{k}_a)
    B_\zeta(\vect{k}_1,\vect{k}_2,\vect{k}_3) ~,
\end{equation}
The strength of the non-Gaussian signal in the bispectrum can conveniently parameterised in terms of the dimensionless non-linearity parameter $f_{NL}$, by defining
\begin{equation}
    \label{f_{NL}}
    B_\zeta \equiv- \frac{6}{5} f_{NL}
    [ P_{\zeta}(k_1) P_{\zeta}(k_2)+ \mbox{2 permutations} ]
    \,.
\end{equation}

Now, from the semiclassical relations, we can also compute the effect of the long wavelength background of scalar modes on the three point functions
\bea
\langle\zeta_{k_1}\zeta_{k_2}\zeta_{k_3}\rangle_\zbar &=& \left[1+\zbar \frac{\partial}{\partial \zbar} + \hf \zbar^2\frac{\partial^2}{\partial\zbar^2}+\cdots\right]\left[e^{-9\zbar}\langle\zeta(e^{-\zbar}k_1)\zeta(e^{-\zbar}k_2)\zeta(e^{-\zbar}k_3)\rangle\right]\\
&=&  \left<\zeta_{k_1}\zeta_{k_2}\zeta_{k_3}\right>_0-(n_{3}-1)\left.\zbar\left<\zeta_{k_1}\zeta_{k_2}\zeta_{k_3}\right>\right|_0+\frac{(n_{3}-1)^2\zbar\zbar}{2}\left.\left<\zeta_{k_1}\zeta_{k_2}\zeta_{k_3}\right>\right|_0+\dots\nonumber
\eea
where like in \cite{Huang:2006eha}, we define 
\beq
n_3-1 = 2(n_s-1)+\frac{d\log(f_{NL})}{d\log(k)}~.
\eeq

Then again completely analogous to the calculation of the effects on the two-point function, we obtain the effect of the soft scalar mode on the three point function, by averaging over the three-point correlation function in the large box
\bea
\left< \left<\zeta_{k_1}\zeta_{k_2}\zeta_{k_3}\right>_{\zbar}\right> &=& \left<\zeta_{k_1}\zeta_{k_2}\zeta_{k_3}\right>_0+\frac{1}{2}(n_{3}-1)^2 \left<\zeta_{k_1}\zeta_{k_2}\zeta_{k_3}\right>_0\left<\zeta^2(x)\right>_*~.
\eea

For simplicity, let us consider the squeezed limit $k_1<<k_2,~k_3$ of single field slow-roll inflation, where at tree-level $f_{NL} =(5/12)(n_s-1)$ \cite{Maldacena:2002vr}.  In this case we then have 
\beq\label{Bcorr}
  B_\zeta = (n_s-1) \left[1+\frac{1}{2}(n_3-1)^2\left<\zeta^2(x)\right>_*)\right]
    [ P^{(0)}_{\zeta}(k_1) P^{(0)}_{\zeta}(k_2)+ \mbox{2 permutations} ]~,
\eeq
where $P^{(0)}_{\zeta}(k)$ is the tree-level power-spectrum. By using the formula for the two point function in eq.~({\ref{sol2}}),
\beq
P_{\zeta}(k)=P^{(0)}_{\zeta}(k)\left[1+\big(\frac{1}{2}(n_s-1)^2+\al_s\big)\left<\zeta^2(x)\right>_*\right]~,
\eeq
we obtain in the squeezed limit
\bea
f_{NL}& =&\frac{5}{12}(n_s-1)\left\{1+\left[\frac{1}{2}(n_3-1)^2-((n_s-1)^2+2\al_s)\right]\left<\zeta^2(x)\right>_*\right\}~,\nonumber\\
&=& \frac{5}{12}(n_s-1)\left[1+((n_s-1)^2-2\al_s)\left<\zeta^2(x)\right>_*\right]~,
\eea
where we in the last line used that in the squeezed limit $f_{NL}$ is independent of $k$. This indicates that at one-loop IR effects shift the magnitude of the non-Gaussianity averaged over the ``large box" upward in single-field slow-roll inflation. We expect tensor modes to also contribute corrections, although in the examples considered in the next section these are relatively suppressed.  We also expect corrections for more general momenta, and higher-order correlators.

\subsection{Example: chaotic inflation}\label{Chaoi}

In chaotic inflation the initial conditions for the inflaton field are determined by large quantum fluctuations  in the ``self-reproduction" regime. In this regime the quantum fluctuations of the inflaton field are larger than the classical evolution of the background field, $\delta\phi > \dot\phi_c \Delta t$  (with $\Delta t\sim 1/H$), and the quantum fluctuations randomly cause some regions to inflate, while in other regions the field value set by the random quantum fluctuations does not allow for inflation to begin.

A simple model is provided by a monomial inflaton potential
\beq\label{monopot}
V(\phi) =\lambda M_p^4\left(\frac{\phi}{M_p}\right)^{\al}~;
\eeq
inflation takes place when $\ep <1$, which implies large field values $\phi> \phi_f =\al /\sqrt{2}$. At very large field values, when the potential energy density becomes of order $\order( M_p^4)$, the low energy effective field theory approach to gravity breaks down. This happens a field values $\phi>\phi_p= \lambda^{-1/\al}$. However, the regular perturbative description of inflaton quantum fluctuations on a classical background evolution only becomes unproblematic at a lower intermediate energy scale, given by the self-reproduction scale,  at field values 
\beq\label{reprob}
\phi<\phi_{i} = (12\pi^2\al^2/\lambda)^{1/(\al+2)}\ . 
\eeq
Thus, the conventional inflationary regime is determined by $\phi_{i}>\phi>\phi_f$. 

In chaotic inflation, the total inflated volume can typically be much larger than the observed universe, if inflation is assumed to begin just at the end of the self-reproduction regime. Therefore, in such a scenario we might expect large IR effects.   We see from eqs.~(\ref{Tscalar}), (\ref{Ttensor}), and (\ref{Bcorr}) that the 
IR effects of the scalar and tensor loops provide fractional corrections given by the quantities
\beq\label{corrdef}
g_{IR} \equiv \big(\frac{1}{2}(n_s-1)^2+\al_s\big) \left< \zeta^2(x)\right>_*~,~ h_{IR}=(n_s-1)  \left< \gamma^2(x)\right>_*\ ,
\eeq
evaluated at the time $t_*$ where the visible modes exit the horizon.
Thus, in order to evaluate their size, we need to evaluate these quantities for the slow-roll potential (\ref{monopot}).

The scalar variance is given in eq.~(\ref{zvar}), and can be computed in terms of the change in the field value and the parameters of the potential.  
The build up of the classical perturbation $\zeta(x)$ on large scales can be regarded as Brownian motion of $\zeta(x)$. The classical perturbation $\zeta(x)$ receives a kick of size $(1/\sqrt{2\ep})H/2\pi$ in a random direction every time interval $\Delta t=1/H$, from the quantum mode $\zeta_{q}$ with $q=aH$, which exits the horizon and becomes classical in the given time interval.
In de Sitter,  with constant $H$, the mean squared distance to the origin is proportional to the time elapsed  \cite{Linde:2005ht}, i.e. $\left<\sigma^2(x)\right> = H^3t/(4\pi^2)$.
In slow-roll, since $H$ are larger at the beginning, the kicks are larger at the beginning, and the variance is expected be larger than $H^3(t) t/(4\pi^2)$. In order to compute the variance of $\zeta$ during inflation, it is therefore convenient to express the integral over $q$ as an integral over field values at time $q= aH$, when the given mode is exiting the horizon.

Specifically, to leading order in slow-roll
\beq
\frac{d\log q}{d\phi} = \frac{d\log q}{d\log a} \frac{d\log a}{dt}\frac{dt}{d\phi}=\frac{H}{\dot \phi} = -\frac{3H^2}{V'} = -\frac{\phi}{\alpha}\ .
\eeq
Then, using $H^2=V/3$ and (\ref{slowparam}) gives
\beq
 \left< \zeta^2(x)\right>_* = -\frac{\lambda}{12\pi^2\alpha^3}\int_{\phi_i}^{\phi_*} d\phi \phi^{\alpha+3} = \frac{\lambda}{12\pi^2\alpha^3(\alpha+4)}(\phi_i^{\alpha+4}-\phi_*^{\alpha+4})\ .
 \eeq
 Note that this is dominated at the upper value, $\phi\approx\phi_i$, so we will neglect $\phi_*$. Next, the condition for the boundary of self-reproduction (\ref{reprob}) gives the value of $\phi_i$ in terms of $\alpha$ and $\lambda$, and thus 
 \beq
 \left< \zeta^2(x)\right>_*\approx \frac{1}{\alpha(\alpha+4)}   (12\pi^2\al^2/\lambda)^{2/(\al+2)}\ ,
 \eeq
  The coupling constant $\lambda$ is then fixed by requiring that the observed value from WMAP 7 \cite{Larson:2010gs}, $\mathcal{P}_{\zeta}(k)= (2.43\pm0.11)\times 10^{-9} $, is obtained. This gives
\beq
\lambda = 12\pi^2\al^2(2\al N_*)^{-1-\al/2}\mathcal{P}_{\zeta}(k)~
\eeq
where $N_*$ is the number of efolds left of inflation when the visible modes exit the horizon,
\beq
N_*=\int_{t_*}^{t_f} H dt \simeq\int_{\phi_f}^{\phi_*} \frac{V}{V'} d \phi = \frac{1}{2\alpha}(\phi_*^2-\phi_f^2)\ .
 \eeq
   Depending slightly on the reheat temperature, we can assume that when the observable modes exit the horizon $N_* = 60$ in order for inflation to have time to solve the horizon and flatness problems of the standard Big Bang model. 
Thus, we rewrite
\beq
 \left< \zeta^2(x)\right>_*\simeq \frac{2N_*}{\alpha+4} \mathcal{P}_{\zeta}(k)^{-2/(\alpha+2)}\ .
 \eeq

We can also express the spectral index $n_s-1=2\eta-6\ep$ in terms of $N_*$, using
\bea
\ep &=& \frac{1}{2}\left(\frac{V'}{V}\right)^2 = \frac{\al^2}{2\phi_*^2}=\frac{\al}{4N_*}\\
\eta &=& \frac{V''}{V} = \frac{\al(\al-1)}{\phi_*^2}=\frac{\al-1}{2N_*}\ .
\eea
These give 
\beq
n_s-1 = -(1+\alpha/2)/N_*\ .
\eeq
Similarly one can compute
\beq
\al_s = -(1+\al/2)/N_*^2\ .
\eeq

Combining these results, we find the fractional size of the scalar contributions,
\beq
 g_{IR}=\big(\frac{1}{2}(n_s-1)^2+\al_s\big) \left< \zeta^2(x)\right>_*\simeq\frac{\alpha^2-4}{4(\alpha+4)} \frac{1}{N_*} \left(\mathcal{P}_{\zeta}(k)\right)^{-1/(1+\al/2)}~.
\eeq

Taking $N_*=60$ and $\mathcal{P}_{\zeta}(k)= (2.43\pm0.11)\times 10^{-9} $  \cite{Larson:2010gs}, the fractional size of the scalar correction parameter is to first order in slow-roll 
\bea
\al &=& 1\qquad \Rightarrow \qquad g_{IR} \simeq -1383\\
\al &=& 2\qquad \Rightarrow \qquad g_{IR} \simeq 0\\
\al &=& 3\qquad \Rightarrow \qquad g_{IR} \simeq 8\\
\al &=& 4\qquad \Rightarrow \qquad g_{IR} \simeq 5
\eea
We see that at first order in slow roll, there is a cancellation for the case $\al=2$, and the leading contribution comes from higher-order in this expansion.

The tensor correction parameter is calculated similarly.  We find
\beq
 \left< \gamma^2(x)\right>_*\ = \frac{\lambda}{6\pi^2\alpha(\alpha+2)}(\phi_i^{\alpha+2}-\phi_*^{\alpha+2})\ ,
 \eeq
 and so
 \beq 
 h_{IR}= (n_s-1)  \left< \gamma^2(x)\right>_*\  \simeq \frac{\alpha}{N_*}\ .
 \eeq

 The tensor contribution is thus suppressed relative to that of the scalar.  For general cases one finds $g_{IR}>1$ and the corresponding perturbative corrections become large before inflation ends, if it is assumed to begin just below the self-reproduction regime. Thus, correspondingly, the corrections to the tensor/scalar ratio and nongaussianity parameters, averaged over the ``large box," are large.

\section{Comments on IR effects}

Infrared divergences in the propagator and variance of a light field have long been known and explored in the inflationary literature.  We have found that 
for long periods of inflation, or pure de Sitter space, these divergences  are associated with large loop corrections to physical quantities in the total inflated volume (``large box").  These can be thought of as due to accumulation of long wavelength fluctuations, that provide significant corrections to correlators, and this phenomenon can be directly analyzed in a semiclassical approach, using so-called ``consistency relations."

One might at first sight be puzzled by the appearance of large logarithms like we have found, given the analysis of \cite{Senatore:2009cf}, which appears to argue that large logarithms don't arise in correct calculations, and in particular are forbidden by the symmetry $a\rightarrow \lambda a$, $x\rightarrow x/\lambda$.  However, the analysis of \cite{Senatore:2009cf} applies to logarithms that are UV large, and not to IR-divergent logs.  Indeed, note that the IR-regulated expressions, {\it e.g.} (\ref{gammaIR}), respect this symmetry.

In a future publication\cite{GiSlinprep} we will explore other manifestations of these and related IR effects.  Such IR effects are clearly physical.  For example, accumulation of IR scalar fluctuations serves as the driving mechanism for self-reproduction.  Here we have shown that a similar buildup of graviton fluctuations leads to corrections to correlators that may be significant.  Such strong IR behavior also enters other physical quantities, such as the tri-spectrum.
 For such IR divergences in modes coupled with gravitational strength, the characteristic time scale at which the corrections become strong is $t\sim RS$, where $R$ and $S$ are the de Sitter radius and entropy, respectively.

The question of how to handle such IR divergences has received considerable discussion.  For example, one might try to resum the IR effects to give a corrected background, and then expand about this background.  As a first approach, consider for simplicity eq.~(\ref{zetaexp}) and eq.~(\ref{sol2}), although the argument will be exactly the same for the tensor modes in eq.~(\ref{zetaIR}) apart from the trivial extra complication of polarization sums. If we continued the Taylor expasion in eq.~(\ref{zetaexp}) to arbitrary order,  eq.~(\ref{sol2}) would become
\bea
\left< \left<\zeta_{k_1}\zeta_{k_2}\right>_{\zbar}\right> &=& \left<\zeta_{k_1}\zeta_{k_2}\right>_0+(2\pi)^3\delta^3(\vect{k}_1+\vect{k}_2)\sum_{n=1}^{\infty}\frac{(2n-1)!!}{(2n)!} \langle\zeta^2(x)\rangle_*^n\left(3+\frac{\p}{\p\log(k)}\right)^{2n}P_{\zeta}(k)\nonumber\\
&=&\left<\zeta_{k_1}\zeta_{k_2}\right>_0\left[1+\sum_{n=1}^{\infty} L_{2n}^{-1}(1+\alpha/2) \left(\frac{n_s-1}{1+\alpha/2}\right)^{2n}
\langle\zeta^2(x)\rangle_*^n\right]\ ,
\eea 
where $L_n^a(x)$ is the generalized Laguerre polynomial, and in the second step we restricted ourselves to monomial potentials of the type eq.~(\ref{monopot}). Interestingly this sum does not converge in general, and the perturbative treatment seems to break down completely
as the variance becomes larger at long distance, and all terms contribute to the sum.  Since $\left<\delta\phi^2(x)\right> \sim H^3t$, we see this happen past $t\sim RS$.  For example, in the model with  $\al=2$ studied in section \ref{Chaoi}, classical evolution led to $(n_s-1)^2\langle\zeta^2(x)\rangle_*=450$.  If we were to take the resummation at face value, it would give a divergent result for $\left< \left< \zeta_{k_1}\zeta_{k_2}\right>_{\zbar}\right> $,
at the end of chaotic inflation, indicating that we had entered a non-perturbative regime\footnote{Note that this behavior is different from the behavior of an $O(N)$ invariant test scalar field in the large $N$ limit, where the exact resummed correlation functions approach a constant for large values of $g_{IR}$ \cite{Riotto}.}, with quantum fluctuations overwhelming the classical background evolution in the ``large box".  While elsewhere it has been advocated\cite{Polyakov:1982ug,Polyakov:2009nq,Polyakov:2007mm,Tsamis:1992sx,Tsamis:1994ca,Tsamis:2007is} that loop effects destabilize de Sitter space by negating the cosmological constant, the physics we apparently encounter here suggests another type of instability of inflationary universes, to growth of large fluctuations of fields and geometry on large scales.  This is similar to self-reproduction, but apparently more general.

While various approaches to handling this regime have been advocated, for example resummation via a dynamical renormalization group approach \cite{Seery:2009hs,Burgess:2009bs} or via stochastic approaches \cite{Starobinsky:1994bd,Tsamis:2005hd, Riotto}, no method has yet been found for a precise and complete treatment of the quantum dynamics into the regime where such effects become important. It is not clear how the large corrections can be absorbed into a background correction. It may be that the regime where such effects become important represents a true boundary in our description of physics via perturbative gravity, and that complete description requires fully non-perturbative gravity, quite possibly with new dynamics.

While some IR divergences are dimension dependent, note that the ones discussed here appear to persist in higher dimensional gravity.  In $D$ spacetime dimensions, dimensional analysis tells us that in a time $1/H$, the fluctuation of a scalar field is $\delta\phi^2\sim H^{D-2}$.  This in turn indicates growth $\left<\delta\phi^2(x)\right> \sim H^{D-1}t$, again becoming large on a time scale $t\sim RS$.

Note also that IR divergences are typical indicators that one has not asked a sensible question, as in the story of soft photon effects in QED and the need to calculate inclusive sums.  
In the regime where IR effects become important in gravity, it may be that there is a good {\it local} description of the dynamics, as in pictures of one inflating bubble of a chaotic universe, described for sufficiently short times.  That is, there may be good local questions, such as the evolution of quantum fields in a limited region of the chaotic universe for a limited time,  and resummation and/or renormalization group methods may be useful for describing such a piece of the universe, where for example the fluctuating backgrounds ({\it e.g.} the scalar vev, and thus vacuum energy in a chaotic scenario) take particular values.\footnote{Moreover, a more basic description of such local questions may be through relational observables, such as the proto-local observables of \cite{Giddings:2005id}.}  
For example, we might view the correlation functions in the large box as the average correlation function in a ``small box" (Hubble sized region) within the large box. This interpretation is manifest the way we calculate our semiclassical relations.  The average correlation function in the large box is not directly measurable for a local observer, and some IR effects might be absorbed into the uncertainty of the state of the fields of the small box in the large-box ensemble; in particular we might not live in an average small box within the large box. This interpretation also resembles some of the ideas in \cite{Lyth:2007jh,Bartolo:2007ti}. In particular, in the special case of single field inflation the end of inflation is determined by the {\it v.e.v.} of the inflaton, and one might condition the probability distribution of the correlation functions on only $60$ e-folds being left of inflation in the small box, using the inflaton as a clock\cite{Unruh:1998ic} (see however also \cite{Unruh:2008zza}). But more global questions, such as that of determining the quantum state of the universe on a global slice, and/or at long times, and associated quantities such as averages over the slice, may not have answers -- at least not in perturbative gravity.

Indeed, this viewpoint that such phenomena could represent a true breakdown requiring new nonperturbative dynamics and mechanisms has been previously suggested  in \cite{Giddings:2007ie,Giddings:2009ae}.  In fact, there is a situation in black hole physics that is quite similar, and is associated with the so-called information paradox.  As originally argued by Hawking\cite{Hawking:1974sw}, perturbative methods imply that black hole evaporation destroys information, but this is very problematic.\footnote{For reviews of the information paradox see \cite{Giddings:1995gd,Strominger:1994tn}.}  A more careful argument requires calculation of the quantum state of fields on a spatial slice spanning both the interior and exterior of the black hole.  Such ``nice slices" are analogous to the spatial slices of de Sitter space or slow-roll\cite{Nimatalk,Giddings:2007ie,ArkaniHamed:2007ky,Giddings:2009ae}.\footnote{Analogies between black holes and de Sitter cosmology have long been pursued; for  other examples see {\it e.g.} \cite{Bousso:2006ge},\cite{ArkaniHamed:2007ky}.}  Note that while in cosmology one may view the state on such a slice as unmeasurable, in the black hole context it is more directly connected to measurable quantities since ultimately we imagine measuring the complete quantum state of  the system ({\it i.e.} all outgoing Hawking radiation). 
Ref.~\cite{Giddings:2007ie,Giddings:2009ae} argued that a perturbative calculation of the state on such a slice breaks down due to accumulated effects of fluctuations, on the time scale $t\sim R S$, where now $R$ and $S$ are the black hole radius and entropy.\footnote{Ref.~\cite{Nimatalk,ArkaniHamed:2007ky} suggested a breakdown of a different origin due to small spacing between nice slices.  Note, though, that they also argue that any attempt to regulate inflation through slow-roll will fail after a time scale $t\sim RS$ as a transition to eternal inflation will result, which does fit with our discussion.  Also, \cite{GiMa} argued that fluctuations of observers/observables become important on this time scale.}  We have demonstrated an analogous growth of perturbative corrections in cosmology.  In the black hole context, \cite{Giddings:2007ie,Giddings:2009ae} proposed  this as a resolution of the information paradox, since such a breakdown means that there is no sharp calculation of the state on the slice, and thus no sharp calculation of the missing information.  While this would eliminate the contradiction producing a paradox, it leaves an information {\it problem}, which is to determine the non-perturbative mechanics that takes over in this regime, and by which the information escapes the black hole, while it is still of a macroscopic size comparable to its original radius.  We see close parallels between this discussion and that of inflationary cosmology.
\vskip.1in
\noindent{\bf Acknowledgements} We wish to thank G. Dvali, J. Hartle, S.F. Hassan, D. Marolf, A. Riotto, and B. Sundborg for discussions. SBG gratefully acknowledges the kind hospitality of the CERN theory group, where this work was initiated. The work of SBG was supported in
part by the U.S. Dept. of Energy under Contract
DE-FG02-91ER40618.

\appendix

\section{Review of consistency relations}

In this appendix we will review the consistency relations for the bispectrum of non-Gaussianity in the squeezed limit \cite{Maldacena:2002vr,Creminelli:2004yq} and for the trispectrum of non-Gaussianity from the exchange of a long wavelength graviton between two pairs of scalar modes in the counter-collinear limit \cite{Seery:2008ax}. 

\subsection{Squeezed limit consistency relation for the 3-point function} 

Consider the three point correlation function of curvature perturbations $\left< \zeta_{k_1} \zeta_{k_2} \zeta_{k_3}\right>$ in the limit where one of the momenta is much smaller than the other two $k_1<<k_2,k_3$. Then the long wavelength mode will rescale the spatial background of the other two modes, as can be seen from the form of the metric 
\beq\label{metr1}
ds^2=-dt^2+e^{2\zeta_1}a^2(t)dx^2~.
\eeq
Now, we can Taylor expand the two-point function of the two short wavelength modes $\zeta_{k_2},\zeta_{k_3}$ on the background of the long wavelength mode $\zeta_1$
\beq
\left<\zeta_{k_2}\zeta_{k_3}\right>_{\zeta_1}= \left<\zeta_{k_2}\zeta_{k_3}\right>_0+\left.\zeta_1\frac{\p}{\p\zeta_1}\left<\zeta_{k_2}\zeta_{k_3}\right>\right|_0+\dots
\eeq
and then by correlation of the variation of the two point function of the shifted background with the long wave mode, one obtains the three point correlation function in the squeezed limit 
\bea\label{cons1}
\lim_{k_1 \to 0} \left< \zeta_{k_1} \zeta_{k_2} \zeta_{k_3}\right>&=& \left< \zeta_{k_1}\left<\zeta_{k_2}\zeta_{k_3}\right>_{\zeta_1}\right> \nonumber\\
& \sim& -(n_s -1) \left<\zeta_{k_1}\zeta_{-k_1}\right>\left<\zeta_{k_2}\zeta_{k_3}\right>~.
\eea

Using the definition of $f_{NL}$ in eq.~(\ref{f_{NL}}), we obtain in this limit from single field slow-roll inflation \cite{Maldacena:2002vr}
\beq
f_{NL} =  \frac{5}{12}(n_s-1)
\eeq

Note, we could just as well have considered the shift in the background due to a long wavelength graviton mode $\gamma^B$, which would have the effect $dx^2 \to dx^2 +\gamma_{ij}^Bdx^i dx^j$ or in Fourier space $k^2 \to k^2- \gamma_{ij}^Bk_ik_j$. So the we would obtain 

\bea
\left<\zeta_{k_2}\zeta_{k_3}\right>_{\gamma_1}&=& \left<\zeta_{k_2}\zeta_{k_3}\right>_0+\left.{\gamma_1}_{ij}\frac{\p}{\p{\gamma_1}_{ij}}\left<\zeta_{k_2}\zeta_{k_3}\right>\right|_0+\dots\nonumber\\
&=&  \left<\zeta_{k_2}\zeta_{k_3}\right>_0-\left.\gamma_{k_1} \ep_{ij}k_ik_j\frac{\p}{\p k^2}\left<\zeta_{k_2}\zeta_{k_3}\right>\right|_0+\dots
\eea
where the $\ep_{ij}$ is the polarization of the Fourier mode $\gamma_{k_B}$. One then obtains 
\bea
\lim_{k_1 \to 0} \left< \gamma_{k_1} \zeta_{k_2} \zeta_{k_3}\right>&=&  \left< \gamma_{k_1}\left<\zeta_{k_2}\zeta_{k_3}\right>_{\gamma_1}\right>=\frac{4-n_s}{2}\frac{\ep_{ij}k_ik_j}{k^2} \left<\gamma_{k_1}\gamma_{-k_1}\right>\left<\zeta_{k_2}\zeta_{k_3}\right>
\eea

In the same way one could calculate 
\bea
\lim_{k_1 \to 0} \left< \gamma_{k_1} \gamma_{k_2} \gamma_{k_3}\right>&=& \left< \gamma_{k_1}\left<\gamma_{k_2}\gamma_{k_3}\right>_{\gamma_1}\right>=\frac{3-n_t}{2}\frac{\ep_{ij}k_ik_j}{k^2} \left<\gamma_{k_1}\gamma_{-k_1}\right>\left<\gamma_{k_2}\gamma_{k_3}\right>
\eea

All of the above relations have been verified by explicit in-in calculations in  \cite{Maldacena:2002vr}.

\subsection{Counter-collinear limit of graviton exchange diagram} 

Consider the contribution to the four point correlation function $\left<\zeta_{k_1}\zeta_{k_2}\zeta_{k_3}\zeta_{k_4}\right>$ from the exchange of a graviton between two pairs of scalar modes in the limit of the momentum of the exchanged graviton going to zero. Defining $\vec{k}_{12}= \vec{k}_1+\vec{k}_2$, this limit corresponds to the so called counter-collinear limit with $k_{12}<< k_1\approx k_2$, $k_3\approx k_4$.

In the limit where the momentum, $k_{12}$, of the exchanged graviton goes to zero, its effect is again to rescale the background $dx^2 \to dx^2 +\gamma_{ij}^Bdx^i dx^j$ or $k^2 \to k^2- \gamma_{ij}^Bk_ik_j$. Taylor expanding again on the background of the long wavelength graviton  
\bea
\left<\zeta_{k_2}\zeta_{k_3}\right>_{\gamma^B}&=& \left<\zeta_{k_2}\zeta_{k_3}\right>_0+\left.\gamma_{ij}^B\frac{\p}{\p\gamma_{ij}^B}\left<\zeta_{k_2}\zeta_{k_3}\right>\right|_0+\dots\nonumber\\
&=&  \left<\zeta_{k_2}\zeta_{k_3}\right>_0+\left.\gamma_{k_B} \ep_{ij}k_ik_j\frac{\p}{\p k^2}\left<\zeta_{k_2}\zeta_{k_3}\right>\right|_0+\dots~,
\eea
one can obtain the semiclassical contribution to the four point function from the correlation between a pair of two point functions due to the long wavelength graviton\cite{Seery:2008ax}
\bea
\lim_{k_{12} \to 0} \left< \zeta_{k_1} \zeta_{k_2} \zeta_{k_3}\zeta_{k_4}\right>&=& \left< \left<\zeta_{k_1}\zeta_{k_2}\right>_{\gamma_{k_{12}}}\left<\zeta_{k_3}\zeta_{k_4}\right>_{\gamma_{k_{12}}}\right>\nonumber\\
&=& \frac{(n_t-3)^2}{4}\frac{\ep_{ij}k_ik_j}{k^2}\frac{\ep_{lk}k_lk_k}{k^2} \left<\zeta_{k_1}\zeta_{k_2}\right>\left<\gamma_{k_{12}}\gamma_{-k_{12}}\right> \left<\zeta_{k_3}\zeta_{k_4}\right>
\eea

This relation has been checked by taking the appropriate limits of the full in-in calculation in \cite{Seery:2008ax}, where the straightforward generalization to scalar exchange diagrams was also discussed.

\section{Detailed derivations of the semiclassical relations}

In this appendix, we give the detailed derivations of the semiclassical relations for the one-loop IR corrections to the two point functions from virtual tensors and scalars in the loop.

\subsection{Details of the derivation for virtual gravitons}

We want to calculate $\left<\zeta_{k_1}\zeta_{k_2}\right>$ on the shifted background of infrared gravitons. In the perturbed spatial metric
\beq
h_{ij} = a^2e^{2\zeta}\left[e^{\gamma}\right]_{ij}\approx a^2e^{2\zeta}\left[\delta_{ij}+\gamma_{ij}+\dots \right]~,
\eeq
the effect of an infrared graviton $\gamma^B$ is therefore to take $dx^2 \to d\tilde x^2= dx^2+\gamma^B_{ij}dx^idx^j+\dots$. Thus, we want to calculate the average correlation function in the large box, given by
\beq
\left<\left<\zeta_{k_1}\zeta_{ k_2}\right>_B\right>~.
\eeq
This is the average in the large box of $\left<\zeta_{ k_1}\zeta_{k_2}\right>_B$, where $\left<\zeta_{k_1}\zeta_{k_2}\right>_B$ is the correlation function of the short wavelength modes in the background of the long wavelength modes somewhere inside the large box. In real space we can then Taylor expand the correlation function on the unperturbed background
\bea
\left<\zeta({x_1})\zeta({x_2})\right>_B&=& \left<\zeta({x_1})\zeta({ x_2})\right>_0+\gamma_{ij}^B( x_0)\frac{\p}{\p \gamma^B_{ij}}\left.\left<\zeta({{\tilde x}_1})\zeta({{\tilde x}_2})\right>\right|_{\gamma_B=0} \nonumber\\
& &+\frac{1}{2}\gamma_{ij}^B( x_0)\gamma_{kl}^B( x_0)\frac{\p}{\p \gamma^B_{ij}}\frac{\p}{\p \gamma^B_{kl}}\left.\left<\zeta({{\tilde x}_1})\zeta({{\tilde x}_2})\right>\right|_{\gamma_B=0}+\dots\ .
\eea

Using that 
\beq
\gamma_{ij}(x)= \int \frac{d^3 k}{(2\pi)^3}\sum_{s=\pm}\ep_{ij}^s(k)\gamma_{\vec k}^s(t) e^{i\vec k\cdot \vec x}~,
\eeq
we can then compute $\left<\zeta_{k_1}\zeta_{k_2}\right>_B$, by a Fourier transformation
\bea \label{qwas}
\left<\zeta_{k_1}\zeta_{k_2}\right>_B &=& \int\int d^3 x_1 d^3 x_2  e^{-i \vec x_1\cdot\vec k_1} e^{-i \vec x_2\cdot\vec k_2}\left<\zeta( x_1)\zeta(x_2)\right>_B\nonumber\\
&=& \left<\zeta_{k_1}\zeta_{k_2}\right>_0 \nonumber\\
& &+ \int\int d^3 x_1 d^3 x_2  e^{-i \vec x_1\cdot\vec k_1} e^{-i \vec x_2\cdot\vec k_2}\nonumber\\
& &   \times\left(\gamma^B_{ij}(x_0)\frac{\p}{\p\gamma^B_{ij}}+\frac{1}{2}\gamma^B_{ij}(x_0)\gamma^B_{lk}(x_0)\frac{\p}{\p\gamma^B_{ij}}\frac{\p}{\p\gamma^B_{kl}} \right)\nonumber\\
& & \times\left.\left[ \int\int \frac{d^3\tilde{q}_1}{(2\pi)^3}\frac{d^3 \tilde q_2}{(2\pi)^3}e^{i\vec{\tilde x}_1 \cdot\vec{ \tilde  q}_1}e^{i\tilde{\vec  x}_2 \cdot \vec{\tilde q}_2}\left<\zeta_{{\tilde q}_1}\zeta_{{\tilde q}_2}\right>_B\right]\right|_{\gamma_B=0}\ .
\eea

Since the long wavelength mode $\gamma_B$ is almost constant over the scales of variation of the short wavelength modes, we can choose $x_0$ freely between $x_1$, $x_2$. For simplicity we can take $\vec x_0 =(\vec x_1+\vec x_2)/2$. 

In order to evaluate the expression above, it is useful to write
\beq
{\tilde x}^2 = \left[ e^{\gamma^B}\right]_{ij}x^i x^j =  x^2+\gamma^B_{ij}x^i x^j+\dots~,
\eeq
and using that momentum transforms inversely, define
\beq
{\tilde q}^2 = \left[ e^{-\gamma^B}\right]^{ij}q_i q_j =  q^2-\gamma^B_{ij}q_i q_j+\dots~,
\eeq
such that $\vec{\tilde x}\cdot \vec{\tilde q} = \vec x\cdot \vec q$, and
\beq
\frac{\p}{\p\gamma_{ij}^B}=\frac{\p\tilde q^2}{\p\gamma_{ij}^B}\frac{\p}{\p\tilde q^2} = -q_iq_j\frac{\p}{\p \tilde q^2}~.
\eeq
Note also $\det [\exp(\gamma^B)]=1$, so the Jacobian of the transformation $\tilde q \to q$ is trivial, and $\delta^3(\tilde{\vec q}_1+\tilde{\vec q}_2) = \delta^3(\vec{q}_1+\vec{q}_2)$, so we can write
\bea
\left<\zeta_{{\tilde q}_1}\zeta_{{\tilde q}_2}\right>_B& =&(2\pi)^3\delta^3(\tilde{\vec q}_1+\tilde{\vec q}_2) \left<\zeta_{{\tilde q}}\zeta_{-{\tilde q}}\right>\nonumber\\
&=&(2\pi)^3\delta^3(\vec{q}_1+\vec{q}_2)\left[1+\left(-\gamma^B_{ij}q_iq_j +\frac{1}{2}\gamma^B_{il}\gamma^B_{lj}q_iq_j+\dots\right)\frac{\p}{\p q^2}\right.\nonumber\\
& &\left.+\frac{1}{2}
\left((\gamma^B_{ij}q_iq_j)^2+\dots\right) \frac{\p}{\p q^2}\frac{\p}{\p q^2}+\dots\right] \left<\zeta_{{ q}}\zeta_{-{ q}}\right>
\eea
Inserting this into eq.~(\ref{qwas}) and changing the integrations over $\tilde q$ into integrations over $q$, we then obtain after taking the average in the large box, and after integrating out three delta functions
\bea
\left<\left<\zeta_{k_1}\zeta_{k_2}\right>_B\right> &=& \left<\zeta_{k_1}\zeta_{k_2}\right>_0\\
& & +\frac{1}{2}(2\pi)^3\delta^3(\vec k_1+\vec k_2)\int\frac{d^3 k_B}{(2\pi)^3}\sum_s \left<\gamma^s_{k_B}\gamma^s_{-k_B}\right>\nonumber\\
& &\times \left[ \ep^s_{il}(k_B)\ep^{s*}(k_B)_{lj}k_ik_j\frac{\p}{\p k^2}+\ep^s(k_B)_{ij}k_ik_j\ep^{s*}(k_B)_{kl}k_k k_l \frac{\p}{\p k^2}\frac{\p}{\p k^2} \right]P_{\zeta}(k)\nonumber
\eea
Inserting the expression  (\ref{polarsum}) for the polarization sums 
\beq
\sum_s \ep^s_{il}(k_B)\ep^{s*}(k_B)_{lj}k_ik_j = 2 k^2\sin^2(\theta)\quad ,\quad
\sum_s \ep^s(k_B)_{ij}k_ik_j\ep^{s*}(k_B)_{kl}k_k k_l = k^4 \sin^4(\theta)~,
\eeq
and then doing the angular integrals, we finally obtain
\beq
\left< \left<\zeta_{k_1}\zeta_{k_2}\right>_B\right> = \left<\zeta_{k_1}\zeta_{k_2}\right>_0+(2\pi)^3\delta^3(\vec k_1 +\vec k_2)\frac{2}{3}\left<\gamma^2(x)\right>_* \left[ \frac{2}{5} k^4\left(\frac{\p}{\p k^2}\right)^2+k^2\frac{\p}{\p k^2}\right] P_{\zeta}(k_1)\ ,
\eeq
in agreement with eq.~(\ref{zetaIR}).

\subsection{Details of derivation for virtual scalars}

In this subsection we want calculate $\left<\left<\zeta_{ k_1}\zeta_{ k_2}\right>_B\right>$ from the shifted background of infrared scalar modes. For convenience let us neglect for a moment the tensor modes and expand the perturbed spatial metric as
\beq
h_{ij} = a^2e^{2\zeta}\delta_{ij}\approx a^2\delta_{ij}(1+2\zeta+\dots)
\eeq
The effect of an infrared mode $\zeta^B$ is therefore to take $x \to \tilde x= x+\zeta^B x+\dots$. 
 
Like in the previous subsection, we Taylor expand the correlation function on the unperturbed background
\bea
\left<\zeta({x_1})\zeta({ x_2})\right>_B&=& \left<\zeta({ x_1})\zeta({ x_2})\right>_0+\zeta_B( x_0)\frac{\p}{\p \zeta_B}\left.\left<\zeta({{\tilde x}_1})\zeta({{\tilde x}_2})\right>\right|_{\zeta_B=0} \nonumber\\
& &+\frac{1}{2}\zeta_B(x_0)\zeta_B( x_0)\frac{\p}{\p \zeta_B}\frac{\p}{\p \zeta_B}\left.\left<\zeta({{\tilde x}_1})\zeta({{\tilde x}_2})\right>\right|_{\zeta_B=0}+\dots
\eea

Using that 
\beq
\zeta(x)= \int \frac{d^3 k}{(2\pi)^3}\zeta_{ k}(t) e^{i\vec k\cdot \vec x}~,
\eeq
we can then compute $\left<\zeta_{k_1}\zeta_{k_2}\right>_B$, by a Fourier transformation
\bea
\left<\left<\zeta_{ k_1}\zeta_{ k_2}\right>_B\right>&=& \int \int d^3 x_1 d^3 x_2 e^{-i\vec{x}_1\cdot\vec{k}_1}e^{-i\vec{x}_2\cdot\vec{k}_2}\left<\zeta( x_1)\zeta(x_2)\right>_B\\
&=& \left<\zeta_{ k_1}\zeta_{k_2}\right>_0+\frac{1}{2}\int \int   d^3 x_1 d^3 x_2e^{-i\vec{x}_1\cdot\vec{k}_1}e^{-i\vec{x}_2\cdot\vec{k}_2}\int\int \frac{d^3 k_B}{(2\pi)^{3}}\frac{d^3 q_B}{(2\pi)^{3}}e^{i\vec{x}_0\cdot(\vec{k}_B+\vec{q}_B)}\zeta(k_B)\zeta(q_B)\nonumber\\
& &\times\frac{\p}{\p \zeta_B}\frac{\p}{\p \zeta_B}\left.\left[\int\int  \frac{d^3 q_1}{(2\pi)^{3}} \frac{d^3 q_2}{(2\pi)^{3}}e^{-6\zeta_B}e^{i\vec{x}_1\cdot\vec{q}_1}e^{i\vec{x}_2\cdot\vec{q}_2} \left<\zeta({e^{-\zeta_B} q_1})\zeta({e^{-\zeta_B} q_2})\right>_B\right]\right|_{\zeta_B=0}\nonumber
\eea

Since the long wavelength mode $\gamma_B$ is almost constant over the scales of variation of the short wavelength modes, we can choose $x_0$ freely between $x_1$, $x_2$. For simplicity we can take $\vec x_0 =(\vec x_1+\vec x_2)/2$. This enables us to do the x-integrals, which yields a couple of $\delta$-functions, which we can then integrate over to obtain
\bea
 \left< \left<\zeta_{k_1}\zeta_{k_2}\right>_B\right> &=&  \left<\zeta_{k_1}\zeta_{k_2}\right>_0+\frac{1}{2}\int \frac{d^3 k_B}{(2\pi)^{3}}\left<\zeta(k_B)\zeta(-k_B)\right> \frac{\p^2}{\p \zeta^2}\left.\left[e^{-6\zeta_B} \left<\zeta({e^{-\zeta_B}k_1})\zeta({e^{-\zeta_B} k_2})\right>_B\right]\right|_{\zeta_B=0}\nonumber\\
& =&\left<\zeta_{k_1}\zeta_{ k_2}\right>_0+\frac{1}{2}(-3-k\frac{\p}{\p k})^2 \left<\zeta_{ k_1 }\zeta_{ k_2}\right>\int \frac{d^3 k_B}{(2\pi)^{3}} \left<\zeta_{ k_B}\zeta_{-k_B}\right>_0~,
\eea
and using the definition $\left<\zeta_{k_1}\zeta_{ k_2}\right> = (2\pi)^3\delta^3(\vec k_1 +\vec k_2)P_{\zeta}(k_1)$, we obtain the relation 
\bea
 \left< \left<\zeta_{k_1}\zeta_{k_2}\right>_B\right> &=& \left<\zeta_{ k_1}\zeta_{k_2}\right>_0+ \frac{1}{2}(2\pi)^3\delta^3(\vec k_1 +\vec k_2)\int \frac{d^3 k_B}{(2\pi)^{3}}P_{\zeta}(k_B)(3+\frac{\p}{\p\log(k)})^2P_{\zeta}(k_1)\nonumber\\
 &=& \left<\zeta_{ k_1}\zeta_{k_2}\right>_0+ (\frac{1}{2}(n_s-1)^2+\al_s)(2\pi)^3\delta^3(\vec k_1 +\vec k_2)P_{\zeta}(k_1)\int \frac{d^3 k_B}{(2\pi)^{3}} P_{\zeta}(k_B)~.
 \eea
Above, working for simplicity to leading order in slow-roll, we have included the contribution from the running of the spectral index $\al_s=\p n_s/\p \log(k)$ from applying $(3+\p/\p\log(k))$ the second time on $(n_s-1) P_{\zeta}(k_1)$. Note that to higher order in slow-roll, other new interesting non-scale invariant terms proportional to $\log(k)$ begins to appear.
 
Now we can use 
\beq
\left<\zeta^2(x)\right>_* = \int_{a_iH_i}^{a_* H_*} \frac{d^3 k_B}{(2\pi)^3}P_{\zeta}(k_B)~,
\eeq
 to obtain to leading order in slow-roll
\bea \label{relz} 
 \left< \left<\zeta_{\vec k_1}\zeta_{\vec k_2}\right>_B\right> &=& \left<\zeta_{\vec k_1}\zeta_{\vec k_2}\right>_0+ (\frac{1}{2}(n_s-1)^2+\al_s)(2\pi)^3\delta^3(\vec k_1 +\vec k_2)P_{\zeta}(k_1) \left<\zeta^2(x)\right>_*
 \eea
in agreement with eq~(\ref{sol2}).

\section{Loop calculation details}

The steps to calculate the quantities $A_k(\eta)$,  $B_k(\eta)$, and $C_k(\eta)$ are outlined in the main text. We do not include the intermediate steps as they result in somewhat unwieldy expressions, which were calculated via Mathematica.  However, a more complete result, expanded in powers of the UV and IR cutoffs, is:
\bea A_k(\eta)+B_k(\eta) &=&\frac{H^2}{2k^3}\frac{H^2}{(2\pi)^2}\Biggl\{
    \left[\frac{4}{3}\log (k)-2 \log (\Lambda_{IR} )-\frac{2}{3} \log
   \left(\frac{1}{\Lambda_{UV}
   }\right)+\frac{101}{90}\right]\nonumber\\  
   &+&(k\eta)^2\left[\frac{2}{15} \log (k) -\frac{2}{5} \log
   (\Lambda_{IR} ) -\frac{2}{3} \log \left(\frac{1}{\Lambda_{UV} }\right)
   +\frac{241 }{450}\right]\nonumber
    \\
    &+&
    (k\eta)^4\left[\frac{4}{15} \log
   \left(\frac{\Lambda_{UV}}{k}\right) 
    +\frac{4}{75}\right]\Biggr\}
   +O\left(\Lambda_{IR}^2, \frac{1}{\Lambda_{UV}^2
   }\right)\ .
     \eea
 The quantity $C_k$ can also be computed:
 \beq
 C_k(\eta)=   \frac{H^2}{3k^4}\frac{H^2}{(2\pi)^2}  \int\frac{dq}{q}\left(- k^3 q^2 \eta ^4-k^3 \eta
   ^2-\frac{5}{2} k q^2 \eta ^2-\frac{5 q^2}{2 k}-3
   k\right)
   \eeq
   which must be $UV$ and $IR$ regulated.
     
 Likewise, we have computed the analogous integrals $A'_k$, $B'_k$, and $C'_k$ arising from the correlator     $\langle{\dot\sigma}_{k_1}{\dot\sigma}_{k_2}\rangle$.  The sum of the former is
 \bea
 A'_k(\eta)+B_k'(\eta)&=&\frac{H^4}{2k^3}
\frac{H^2}{(2\pi)^2} \Biggl\{(k\eta)^4 \left[\frac{2}{15}  \log (\Lambda_{IR}
   )-\frac{2}{3}  \log \left(\frac{1}{\Lambda_{UV}
   }\right)-\frac{7}{450} -\frac{4}{5}  \log (k)\right]\nonumber\\
  &+&(k\eta)^6
   \left[\frac{4}{15}  \log \left(\frac{k}{\Lambda_{UV}
   }\right)+\frac{38 }{225}-\frac{4k}{15\Lambda_{UV}}\right]\Biggr\}
   +O\left(\Lambda_{IR}
   ^2,\frac{1}{\Lambda_{UV}^2 }\right)    
\eea
and the latter is
\beq
C'_k(\eta) =  \frac{H^4}{3k^4}\frac{H^2}{(2\pi)^2}  \int\frac{dq}{q} \left( k^5 q^2 \eta ^6+ k^5 \eta
   ^4-\frac{5}{2}  k^3 q^2 \eta ^4\right)\ .
   \eeq

\end{document}